\documentclass[twocolumn]{aastex62}

\newcommand{\RNum}[1]{\uppercase\expandafter{\romannumeral #1\relax}}
\newcommand{\bef}{\begin{figure}}      
\newcommand{\eef}{\end{figure}}      
\newcommand{\bea}{\begin{eqnarray}}    
\newcommand{\eea}{\end{eqnarray}}      
\newcommand{\be}{\begin{equation}}      
\newcommand{\ee}{\end{equation}}  
\newcommand\HI{$\textrm{H}\scriptstyle\mathrm{I}$}

\graphicspath{{./}{figures/}}

\received{???}
\revised{???}
\accepted{???}
\shorttitle{Kinematic and dynamics of the galaxy ESO 358-60}
\shortauthors{Sylos Labini \& Straccamore}
\usepackage{amsmath}

\begin{document}

\title{Kinematic and dynamics of the galaxy ESO 358-60}
\author{Francesco  Sylos Labini}
\affil{Centro  Ricerche Enrico Fermi, Via Pansiperna 89a, 00184 Rome, Italy}
\affil{Istituto Nazionale Fisica Nucleare, Unit\`a Roma 1, Dipartimento di Fisica, Universit\'a di Roma ``Sapienza'', 00185 Rome, Italy}
\author{Matteo Straccamore}
\affil{Centro  Ricerche Enrico Fermi, Via Pansiperna 89a, 00184 Rome, Italy}
\affil{Sony CSL - Rome, Joint Initiative CREF-SONY, Centro Ricerche Enrico Fermi, Via Panisperna 89/A, 00184, Rome, Italy}

\correspondingauthor{FSL}
\email{sylos@cref.it}

\begin{abstract}
We investigate the velocity field derived from \HI\; measurements of the irregular galaxy ESO 358-60 using the Velocity Ring Model (VRM) method. This technique, which assumes a coplanar disk, allows us to reconstruct coarse-grained maps of both radial and tangential velocity components from the observed line-of-sight velocity field. {  Such maps reveal} that tangential motions dominate the inner regions, while radial motions become increasingly significant toward the outskirts. This kinematic behavior contrasts with that inferred from the Tilted Ring Model (TRM), which suggests that radial motions are more prominent in the intermediate disk and negligible in the outskirts {  and  detects} a pronounced warp of approximately $20^\circ$, with the inner disk nearly edge-on and the outer regions inclined by approximately $60^\circ$. In contrast, the VRM analysis {  finds}  that the disk exhibits a bar-like structure in its central regions. This interpretation is further supported by the intensity and velocity dispersion maps. To test the origin of the TRM-derived warp, we construct a toy model based on the TRM results and analyze it with the VRM technique, finding evidence that the warp is likely an artifact arising from the TRM's assumptions. Finally, we estimate the galaxy's mass using both the standard dark matter halo model and a dark matter disk (DMD) model, where all mass lies in the disk plane. The DMD yields a total mass approximately three times lower and provides a slightly better fit to the rotation curve.
\end{abstract}

%%%%%%%%%%%%%%%%%%%%%%%%%%%%%%%%%%%%%%%%

\section{Introduction}
The study of the kinematics of external galaxies relies on observations of line-of-sight (LOS) velocity field maps. These maps often reveal more complex patterns than those expected from a purely rotating disk, in which emitters follow circular, stationary, and coplanar orbits around the galactic center. In such an idealized case, the velocity field would appear symmetric with respect to the projected major axis of the galaxy. However, observations frequently show that this symmetry is broken: the kinematic axis, defined as the axis of steepest LOS velocity gradient, {often} varies with radius rather than remaining aligned with the major axis (see, e.g. \cite{Warner_etal_1973,Jorsater+vanMoorsel_1995,Schoenmakers_etal_1997,Zurita_etal_2004,Trachternach_etal_2008,DiTeodoro+Peek_2021}).

In the literature, two broad classes of methods have been introduced for reconstructing galaxy velocity fields from LOS velocity maps. The first class is based on the assumption that the velocities of the emitters are purely circular, with radial motions assumed to be zero. In this framework, deviations from the velocity pattern expected for a purely rotating, coplanar disk are interpreted as arising from variations in the orientation of the orbits as a function of radius. This approach forms the basis of the  \emph{Tilted Ring Model} (TRM), in which the galactic disk is described as a sequence of concentric rings, each characterized by its own orientation, typically parameterized by two free angles: the inclination angle and the position angle (P.A.)\citep{Warner_etal_1973,Rogstad_etal_1974}. While this model can, in principle, incorporate non-zero radial motions, these components must be added only after determining the geometric parameters of the rings. However,  an intrinsic degeneracy between radial motions and the assumed disk geometry prevents a robust assessment of how its underlying assumptions affect the results. As a result, this can lead to a systematic underestimation of the amplitude of radial motions \citep{Fraternali_etal_2001,Schmidt_etal_2016,DiTeodoro+Peek_2021,Wang+Lilly_2023}.

The second class of methods assumes that emitters move on coplanar orbits (i.e., that the disk is flat --- see e.g., \cite{Barnes+Sellwood_2003,Spekkens+Sellwood_2007,Sellwood+Spekkens_2015,Sellwood_etal_2021}) and that their motion includes both tangential and radial components. {  Among these we find } the \emph{Velocity Ring Model} (VRM) {  that} assumes a global inclination angle and P.A. for the disk, and reconstructs, for each radial ring, not only the tangential velocity component \(v_t\) but also the radial component \(v_r\) \citep{SylosLabini_etal_2023b}. A generalization of this model, in which each ring is divided into angular sectors, allows for the reconstruction of \(v_t\) and \(v_r\) within each sector, thereby providing a more detailed characterization of nonaxisymmetric motions and enabling the reconstruction of coarse-grained maps of both velocity components. These maps can be used to study anisotropic structures in the velocity field and correlate them with intensity and velocity dispersion maps, potentially identifying spatial features such as spiral arms, bars, or satellite interactions \citep{SylosLabini_etal_2024,SylosLabini_etal_2025a}.

A key question concerns the validity of the underlying assumption of a flat disk. This can be tested by combining the TRM and VRM methods \citep{SylosLabini_etal_2025b}. Specifically, one can use the radial dependence of the circular velocity and the orientation angles obtained from the TRM to construct a toy disk model, which is then analyzed using the VRM. It can be shown that the signature of a genuinely warped disk, when examined with the VRM, corresponds to a dipolar angular correlation between the radial and tangential velocity components. {  Conversely, the absence of such a dipolar correlation provides a clear diagnostic that a warp can be ruled out.}

The core idea behind this procedure is to use, for {  a given} galaxy, a toy model constructed from the TMR-derived  velocity profiles and geometric deformations as a null hypothesis test. In this way, we can isolate features in the velocity maps that cannot be explained solely by the presence of a warp. These features, therefore, correspond to extrinsic velocity perturbations.

Moreover, such an angular modulation manifests itself as a smooth and gradual variation of both velocity components with polar angle, a pattern that can be accurately detected through VRM analysis. The presence of warps becomes thus evident when studying the velocity fields reconstructed by the VRM: the spatial characteristics of velocity anisotropies, derived under the assumption of a flat disk, offer a clear and distinctive signature of a potential warp. Therefore, if the disk is intrinsically warped, applying the VRM enables us to break the degeneracy between geometric distortions of the disk and genuine radial motions, allowing us to quantify the amplitude of velocity perturbations.

This methodological framework can be directly tested on observational data. We thus consider {  the irregular galaxy}  ESO 358-60, for which a detailed TRM analysis has already been performed.
\cite{Kamphuis_etal_2025} presented a detailed study of the velocity field {  of  ESO 358-60}  with the aim of understanding the morphology and kinematics of its \HI\; disk, and thereby determining whether the galaxy is a member of the Fornax cluster, which lies in the same region of the sky \citep{Serra_etal_2023}. ESO 358-60 is characterized by a quiescent environment compared to other nearby galaxies and shows no clear signs of interaction with a companion. Combined with the fact that the galaxy's systemic velocity lies at the edge of the Fornax velocity distribution, the authors propose that the undisturbed nature of the disk may indicate that ESO 358-60 is not actually a member of the cluster.

Nevertheless, \cite{Kamphuis_etal_2025} modeling using the TRM reveals that, although the \HI\; disk is remarkably regular, the galaxy exhibits a significant line-of-sight warp: the inclination angle varies by approximately \( \sim 20^\circ \) across the disk, with a P.A. variation of about \( \sim 10^\circ \). In addition, they conclude that the disk contains radial motions on the order of 10~km\,s$^{-1}$, extending across the entire optical disk. The best-fit rotation curve shows a flat velocity profile. This profile is measured over a limited radial range, up to \( r \leq 9 \) kpc, and has an amplitude of \( 48.1 \pm 1.4 \) km\,s$^{-1}$. Finally, they do not propose a specific dynamical explanation for the presence of the warp or for the behavior of the radial motions.

Given this context, we have reanalyzed the velocity field of ESO 358-60 using the VRM method. We also performed a consistency test, combining the TRM and VRM approaches, to assess whether the warp identified in the previous analysis is a robust feature. Finally, we estimated the galaxy's mass by applying two different mass models: the Dark Matter Disk (DMD) model \citep{SylosLabini_etal_2023b, SylosLabini_etal_2024, SylosLabini_etal_2025a} and the Navarro-Frenk-White (NFW) halo model \citep{Navarro_etal_1997}.

Both models assume a substantial amount of dark matter to explain the observed rotation curves, as luminous matter alone cannot account for the relatively high velocities observed. However, the spatial and kinematic properties of dark matter are fundamentally different in the two models. In the DMD framework, dark matter is distributed like the luminous matter, sharing the same spatial and kinematic properties. In contrast, the NFW model assumes a quasi-spherical distribution of dark matter, which is thus not flattened as the observed luminous component, and is characterized by a quasi-isotropic velocity dispersion, meaning it is not rotationally supported.

The paper is organized as follows. In Sect.~\ref{sec:data} we describe the observational data. Section~\ref{sec:VRM} presents the results obtained with the VRM method, together with a kinematic and morphological characterization of the galaxy, including an examination of the warp test derived from the TRM analysis. {  In Sect.~\ref{sec:mass_est} we present the mass estimates obtained using the two alternative mass models and in Sect.\ref{sect:orgin} we discuss the physical motivation underlying the DMD framework and its main features.} Finally, Sect.~\ref{sec:dis+con} summarizes and discusses our main conclusions.

%%%%%%%%%%%%%%%%%%%%%%%%%%%%%%%%%%%%%%%%%%%%%%%%%%%%%%%
%%%%%%%%%%%%%%%%%%%%%%%%%%%%%%%%%%%%%%%%%%%%%%%%%%%%%%%
%%%%%%%%%%%%%%%%%%%%%%%%%%%%%%%%%%%%%%%%%%%%%%%%%%%%%%%
%%%%%%%%%%%%%%%%%%%%%%%%%%%%%%%%%%%%%%%%%%%%%%%%%%%%%%%

\section{Data} 
\label{sec:data} 

\begin{figure*}
\centering
%\subfigure[]{\includegraphics[width=0.35\textwidth]{../DATA_11/fig_e_ESO_358-60_M0.pdf.jpg}}
%\subfigure[]{\includegraphics[width=0.35\textwidth]{../DATA_11/fig_e_ESO_358-60_sigma.pdf.jpg}}
\includegraphics[width=0.31\textwidth]{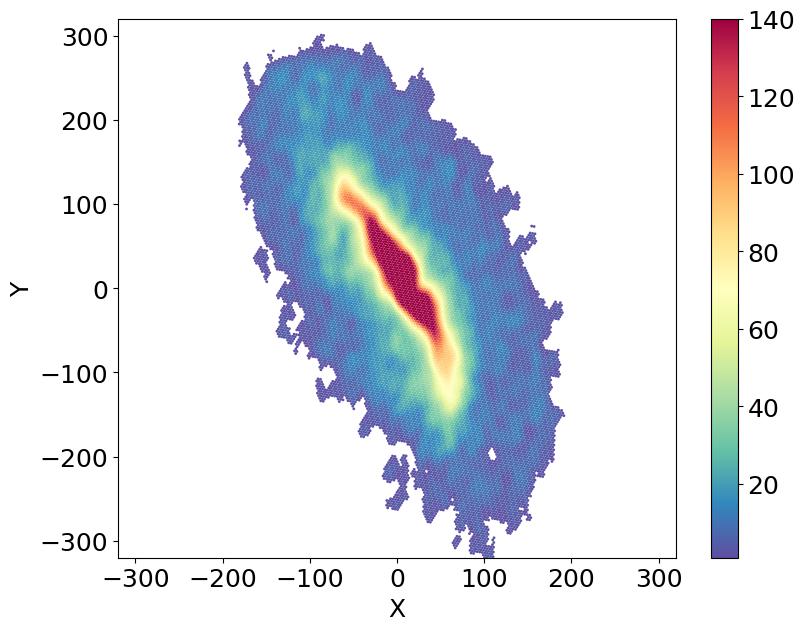}
\includegraphics[width=0.31\textwidth]{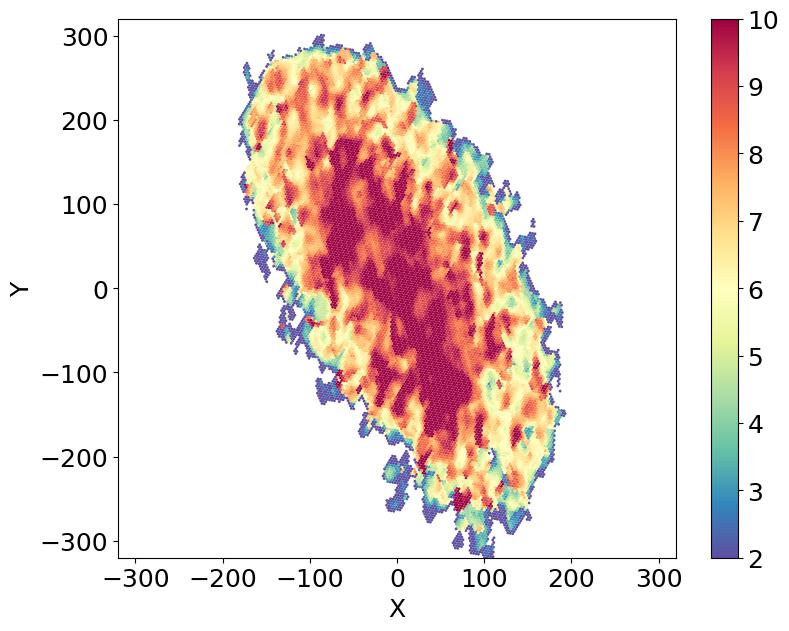}
\includegraphics[width=0.25\textwidth,height=0.25\textwidth]{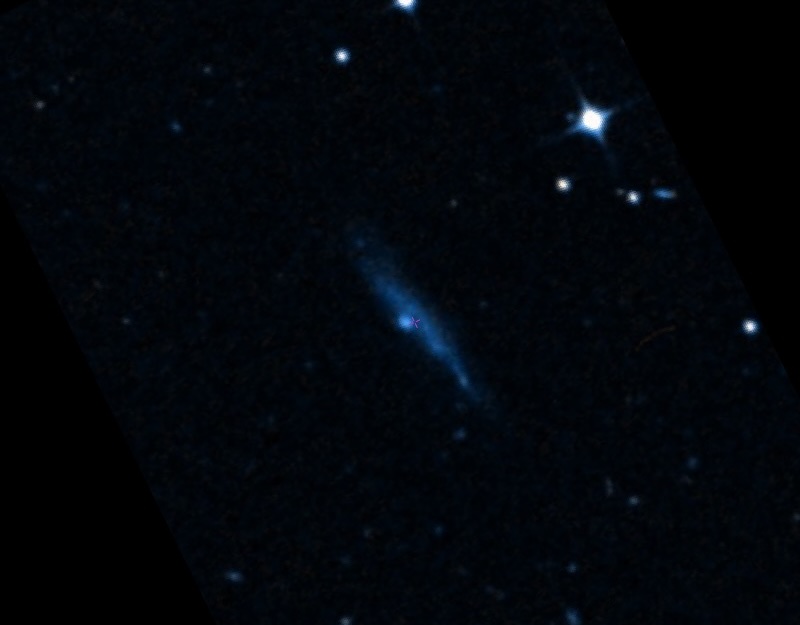}
\caption{
Left panel: neutral hydrogen surface brightness map (moment~0). The color scale indicates the column density in units of $10^{2}\,\mathrm{cm}^{-2}$.  
Middle panel: velocity dispersion map (moment~2). The color scale is given in $\mathrm{km\,s^{-1}}$.  
In both the left and middle panels, the $X$ and $Y$ axes (the coordinates are in the plane of the sky) are in arc-seconds.  
Right panel: Digitized Sky Survey (DSS) image from the HyperLeda galaxy catalog \citep{Makarov_etal_2014} 
 displayed as a color composite. The field of view is about 600 arc-seconds as the neutral hydrogen intensity map.
}
\label{fig:M0M2}
\end{figure*}

The irregular galaxy ESO 358-60 has been observed as part of the \textit{Fornax Survey} \citep{Serra_etal_2023}, which uses \textit{MeerKAT} to map its emission \HI\; 21 cm and investigate its gas distribution and kinematics. The \textit{Fornax cluster}, one of the closest galaxy clusters at a distance of \( D \approx 20  \) Mpc, provides an excellent laboratory for studying  environmental effects on the neutral hydrogen content of cluster galaxies. Fornax is characterized by a high central surface density of galaxies but relatively low total mass, possibly representing a {transitional environment} between massive galaxy groups and rich clusters.

The galaxy ESO 358-60, however, stands out because of its relatively large, regular, and undisturbed \HI\; disk. Because its systemic velocity lies at the edge of the Fornax cluster's velocity distribution, one plausible explanation for the absence of tidal features or an \HI\;  tail is that the galaxy may not be a true cluster member. This hypothesis motivated \citet{Kamphuis_etal_2025} to perform a detailed kinematic analysis. {  They found } that the \HI\; disk is significantly more extended than the observed stellar component, making it difficult to reconcile such an apparently unaffected disk with the galaxy traversing the intra-cluster medium of the Fornax cluster at a velocity of $\gtrsim 600$ km,s$^{-1}$.

Specifically, {  the \HI\; data from the \textit{MeerKAT} survey } were used to perform a kinematic analysis based on the highest resolution data cube available for this galaxy. The high resolution cube has a spatial resolution of $12.2'' \times 9.6''$ (FWHM), a velocity resolution of 1.4 km s$^{-1}$, and a noise level of 0.24 mJy beam$^{-1}$.

The basic properties of ESO~358-60 are listed in Table~1 of \citet{Kamphuis_etal_2025}.  
For our analysis, we adopted a systemic velocity of $806.8 \pm 0.8 \,\mathrm{km\,s^{-1}}$ and a distance of $9.4 \pm 2.5 \,\mathrm{Mpc}$.  
As discussed below, the inclination angle was fixed at $i = 60^\circ$, corresponding to the value determined for the outer regions by \citet{Kamphuis_etal_2025}.  
In contrast, the {  P.A.}  does not play a role in our analysis and is therefore not explicitly considered, {  as it corresponds to an overall rotation of the galaxy}.

The central surface brightness distribution (moment 0) is highly elongated (see  upper panel of Fig.~\ref{fig:M0M2}), suggesting either an edge-on disk or a \textit{bar-like structure} in the inner region. In contrast, the outer regions of the disk appear significantly more circular. Additionally, the surface brightness distributions on the approaching and receding sides are largely symmetric. 

The {  middle panel}  of Fig.~\ref{fig:M0M2} shows the velocity dispersion map (moment 2). Visual inspection reveals a clear correlation: regions of higher intensity tend to exhibit higher velocity dispersion as well.  We will quantify this correlation below.

{ 
 Finally, the Digitized Sky Survey (DSS) image of ESO~358-60, obtained from the HyperLeda galaxy catalog \citep{Makarov_etal_2014}, is shown in the {  right panel} of Fig.~\ref{fig:M0M2}.  The stellar component traced by the DSS image corresponds closely to the neutral hydrogen over-density in the inner region of the galaxy.  
}

%%%%%%%%%%%%%%%%%%%%%%%%%%%%%%%%%%%%%%%%%%%%%%%%%%%%%%%
%%%%%%%%%%%%%%%%%%%%%%%%%%%%%%%%%%%%%%%%%%%%%%%%%%%%%%%
%%%%%%%%%%%%%%%%%%%%%%%%%%%%%%%%%%%%%%%%%%%%%%%%%%%%%%%

\section{Kinematics and morphology} 
\label{sec:VRM} 

%%%%%%%%%%%%%%%%%%%%%%%%%%%%%%%%%%%

\subsection{Tilted Ring Model} 

{  The  TRM} analysis performed by \citet{Kamphuis_etal_2025} led to the conclusion that the \HI\;\ disk of ESO~358-60 is remarkably regular. However, the modeling also revealed the presence of a \textit{significant line-of-sight warp} and radial motions of approximately $10$~km\,s$^{-1}$ extending across the entire optical disk. To find these results they have applied a two-step fitting procedure.

An initial TRM fit was obtained by applying {\tt pyFAT} \citep{Kamphuis_etal_2015,Kamphuis_2024} to the data cube. Since {\tt pyFAT} does not account for radial motions, these must be added manually. For this reason, the TRM results are used as a starting point and are refined through guided fitting with {\tt TiRiFiC} \citep{Jozsa_etal_2007}. Once the outer warp is satisfactorily reproduced in the model, radial motions are introduced to further improve the fit.
The best-fit TRM parameters from \citet{Kamphuis_etal_2025} can be summarized as follows:
\begin{itemize}
  \item The inclination angle is approximately $i \approx 80^\circ$ up to $R \sim 80''$, then decreases linearly to $i \approx 60^\circ$ at $R \sim 150''$, remaining constant out to the outermost regions, i.e. $R \sim 200''$ (see Fig.5 panel (c) of \citet{Kamphuis_etal_2025}).
  \item {  The P.A. is} $\approx 100^\circ$ up to $R \sim 100''$, beyond which it increases linearly, reaching $\sim 110^\circ$ in the outer disk  (see Fig.5 panel (d) of \citet{Kamphuis_etal_2025}).
\end{itemize}

These variations imply a pronounced warp of $\sim 20^\circ$ in the intermediate regions --- driven by the change in inclination --- and a more moderate warp in the outermost parts of the disk, due to the $\sim 10^\circ$ variation in P.A.

The rotation velocity increases approximately linearly at small radii, reaching a plateau of $\approx 48$~km\,s$^{-1}$ at $R \sim 160''$, and remains flat beyond that  (see Fig.5 panel (a) of \citet{Kamphuis_etal_2025}).

The radial velocity profile displays a non-monotonic structure: starting at $\approx 10$~km\,s$^{-1}$ near the center, it decreases to a minimum of $\approx 4$~km\,s$^{-1}$ at $R \sim 30''$, then rises to a maximum of $\approx 15$~km\,s$^{-1}$ at $R \sim 100''$, and finally declines, becoming negative in the outermost regions (see Fig.5 panel (e) of \citet{Kamphuis_etal_2025}).

\subsection{Surface brightness and velocity dispersion profiles} 

\begin{figure*}
\centering
%\subfigure[]{\includegraphics[width=0.35\textwidth]{../DATA_11/fig_e_ESO_358-60_M0.jpg}}
%\subfigure[]{\includegraphics[width=0.35\textwidth]{../DATA_11/fig_e_ESO_358-60_M2.jpg}}
\includegraphics[width=0.45\textwidth]{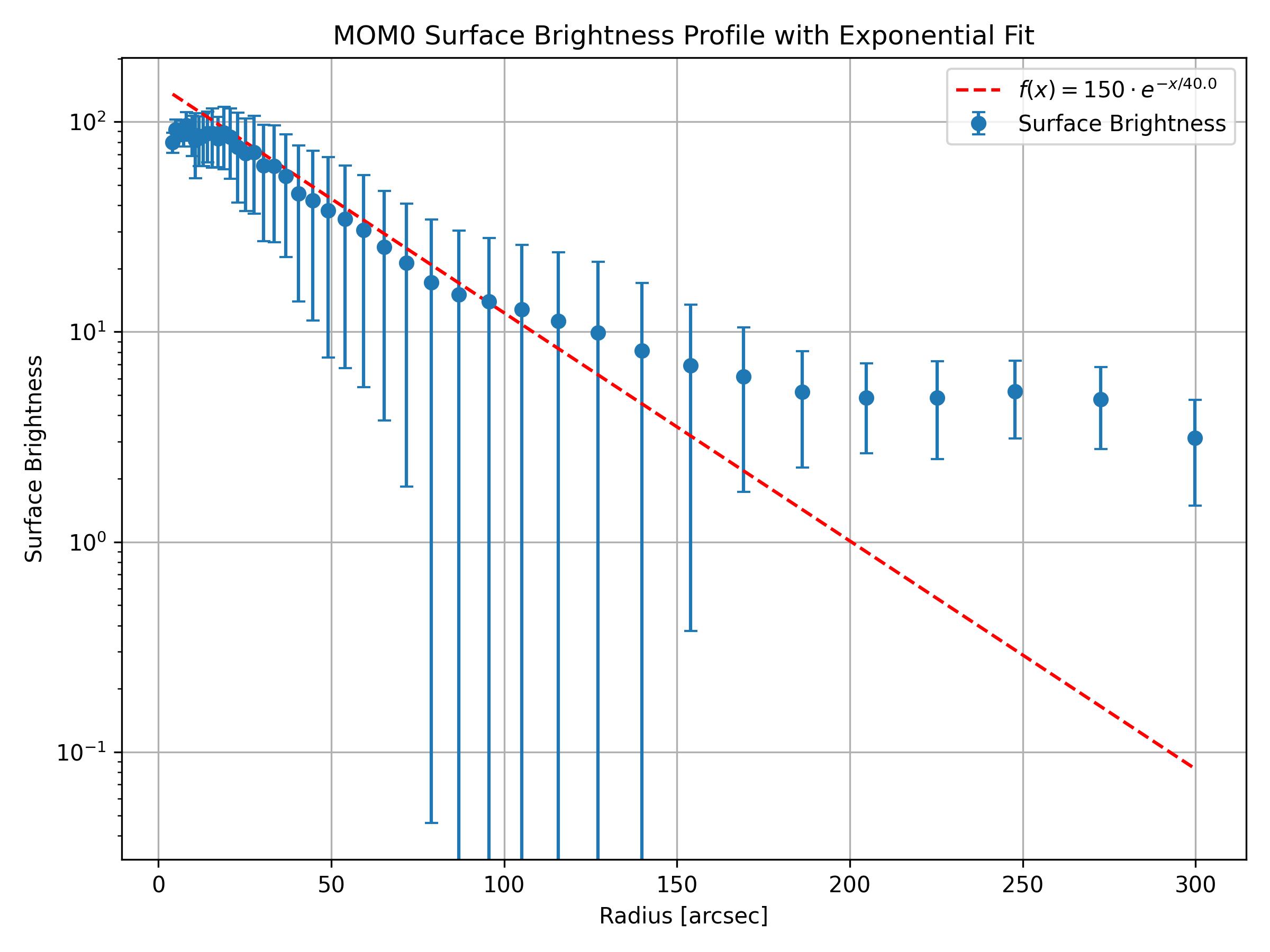}
\includegraphics[width=0.45\textwidth]{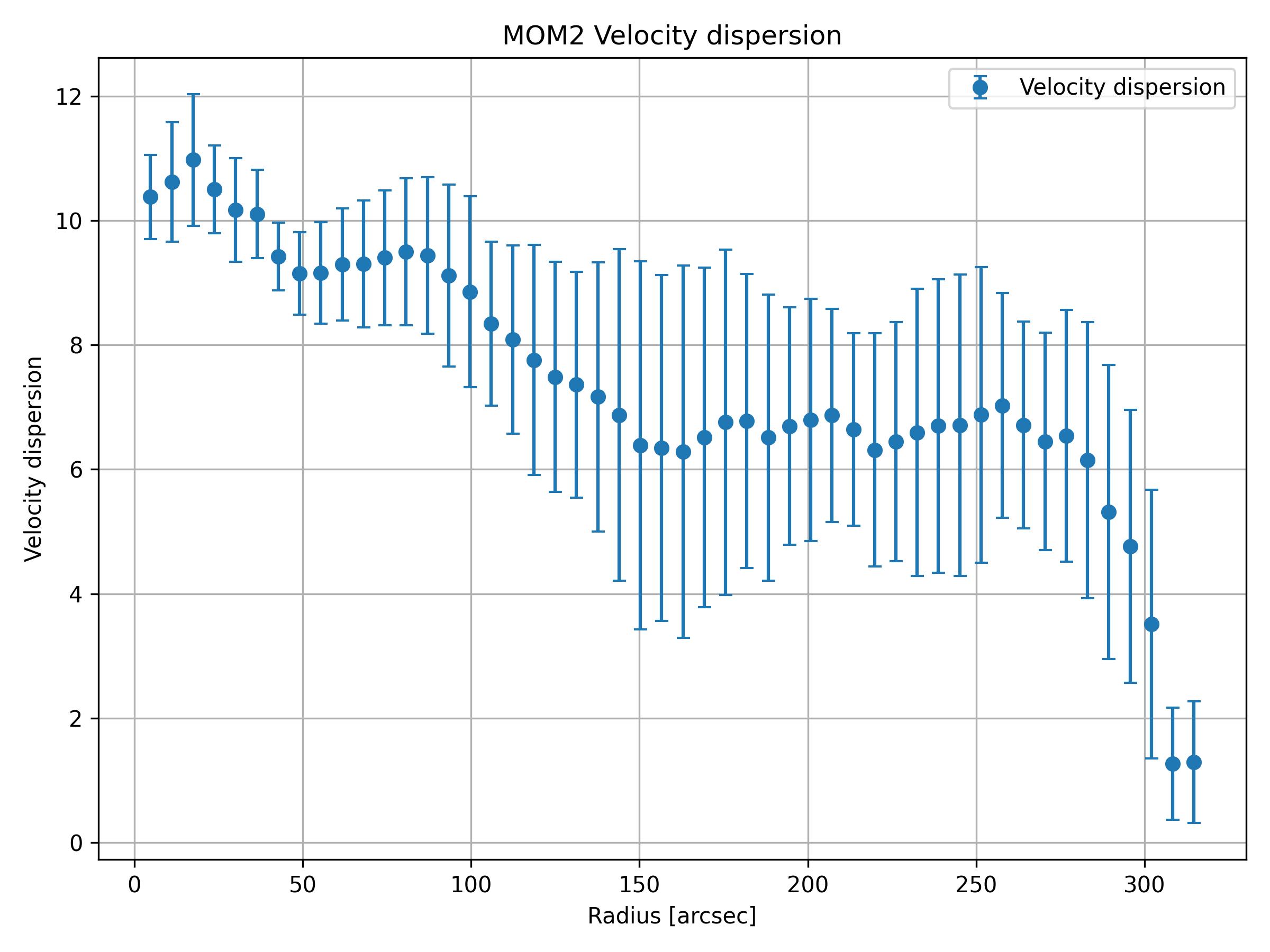}
\caption{
Left panel:   radial profile of the surface brightness.
{  For comparison, a reference line with an exponential decay of characteristic length $40''$ is also shown.  }
Right panel: radial profile of the velocity dispersion.
{  The error bars are determined from the angular average within each ring.  }
} 
\label{fig:prof_M0M2} 
\end{figure*}

We assume that the disk is flat and has an inclination angle of $i = 60^\circ$, consistent with the value estimated by the TRM in the outer regions of the disk \citep{Kamphuis_etal_2025}. Under this assumption, Fig.~\ref{fig:prof_M0M2} shows the profiles of the surface brightness (moment 0) and the velocity dispersion (moment 2), computed in concentric rings. The radial coordinate used corresponds to the de-projected radius in the plane of the galaxy \footnote{We refer  to \cite{SylosLabini_etal_2023b} for a detailed discussion on the transformation from sky-plane polar coordinates $(r, \phi)$ to disk-plane coordinates $(R, \theta)$, in the case of constant inclination angle and P.A. across the disk.}.

The surface brightness profile exhibits an approximately exponential decay, with a characteristic scale length of $R_c \approx 40''$ up to $\sim 150''$, which corresponds to the extent of the elongated central density enhancement. Beyond this radius, the decay becomes significantly shallower. The velocity dispersion profile displays a similar trend: a relatively rapid decline (in a linear-linear plot) within the central $\sim 150''$ radius, followed by a slower decrease in the outer regions. This behavior suggests a correlation between velocity dispersion and surface brightness. A more quantitative analysis of this correlation is presented in the following sections.

\subsection{Velocity ring model} 

The VRM is a method that, under the assumption of a flat galactic disk, allows the reconstruction of two-dimensional maps of the transverse and radial velocity components from observations of the LOS velocity field, \( v_{\mathrm{los}}(r, \phi) \), as a function of the angular coordinates on the plane of the sky in the projected image of an external galaxy \citep{SylosLabini_etal_2023b}. In this framework, the LOS component of the velocity can be expressed as
\begin{equation}
\label{vlos_eq} 
v_{\mathrm{los}} = \left[ v_t \cos(R, \theta) + v_r \sin(R, \theta) \right] \sin(i),
\end{equation}
where \( (R, \theta) \) are the polar coordinates in the plane of the galaxy.

The galactic disk is divided into \( N_r \) concentric rings, each assumed to share the same inclination angle and P.A. The velocity field within each ring is decomposed into a radial component \( v_r = v_r(R) \) and a transverse component \( v_t = v_t(R) \). By inverting Eq.~\ref{vlos_eq}, the VRM yields the values of \( v_r(R) \) and \( v_t(R) \), corresponding to the average profiles of the two velocity components in each ring.

A more refined analysis consists of subdividing each ring into \( N_a \) angular sectors (arcs), resulting in a total of \( N_{\mathrm{cells}} = N_r \times N_a \) cells. The VRM then reconstructs the two velocity components \( v_r = v_r(R,\theta) \)  and \( v_t = v_t(R,\theta) \) in each of the \( N_{\mathrm{cells}} \), yielding a total of \( 2 \times N_{\mathrm{cells}} \) free parameters.

\begin{figure*}
\centering
\includegraphics[width=0.45\textwidth]{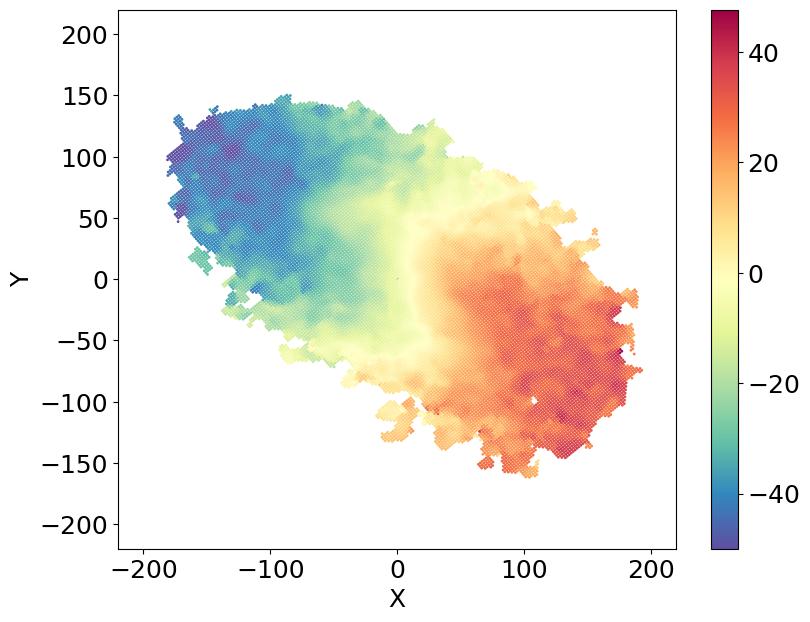}
\includegraphics[width=0.45\textwidth]{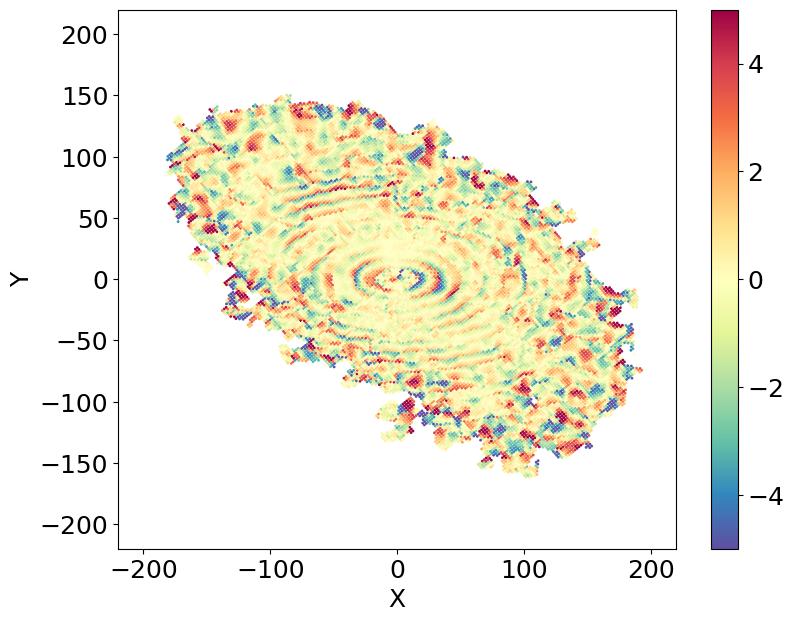}
\caption{Left panel: 
the observed 2D line-of-sight velocity field.
Right panel:  the residuals from the fit to the VRM model with $N_r=20$ and $N_a=32$.
{  The $X$ and $Y$ axes (the coordinates are in the plane of the sky) are in arc-seconds.  }
} 
\label{vlos_fig} 
\end{figure*}

Fig.~\ref{vlos_fig} shows the observed 2D line-of-sight velocity field (left) and the residuals from the VRM fit with $N_r = 20$ and $N_a = 32$ (right). The image is rotated relative to Fig.~\ref{fig:M0M2} to align the zero-velocity line with the $y$ axis, following the convention adopted in the VRM analysis \citep{SylosLabini_etal_2023b}.
{  The kinematic axis} is not straight and the observed image is curved. As mentioned above, the TRM interprets such a curvature as a non-flat, i.e. warped, disk model. In the VRM framework, this curvature is instead related to more consistent radial velocities than those estimated by \citep{Kamphuis_etal_2025}, as we are going to show below. The residual velocity field has a very small amplitude, consistent with the excellent fit provided by the VRM.

\begin{figure*}
\centering
\includegraphics[width=0.22\textwidth]{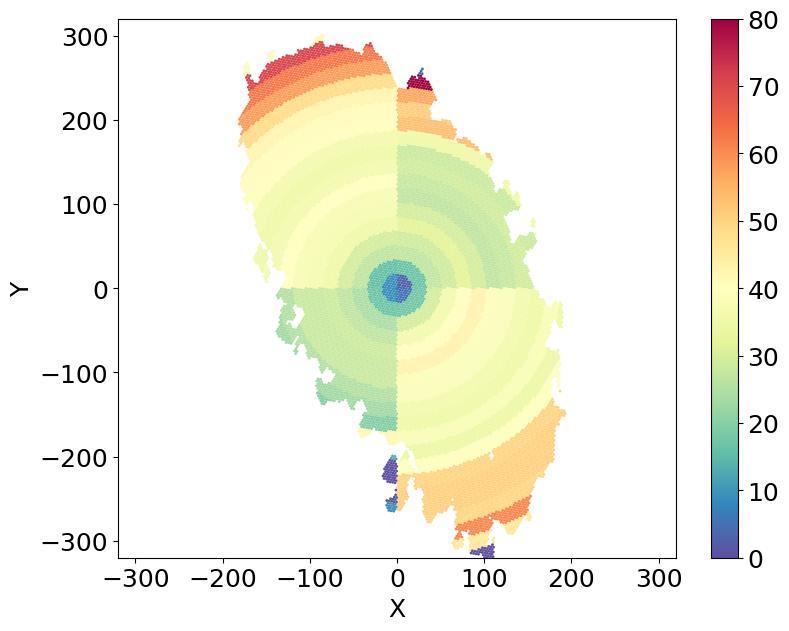}
\includegraphics[width=0.22\textwidth]{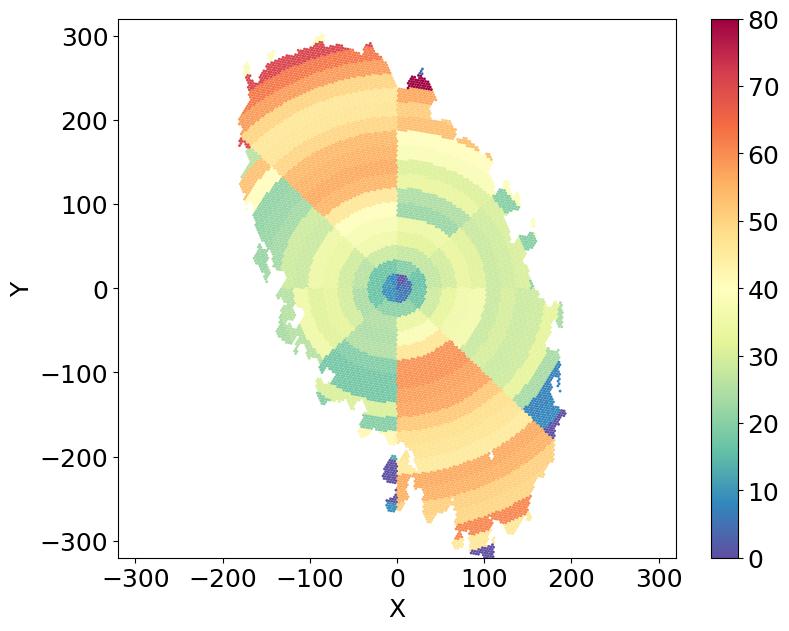}
\includegraphics[width=0.22\textwidth]{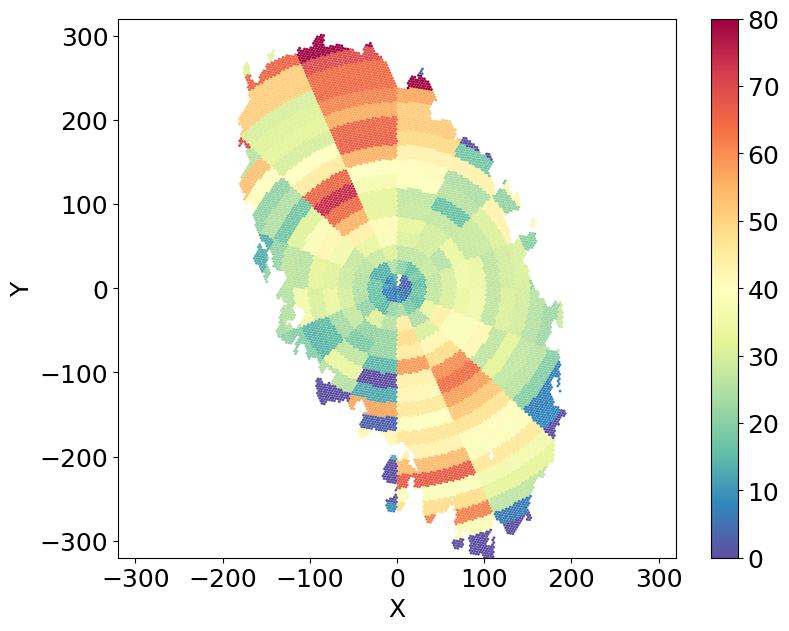}
\includegraphics[width=0.22\textwidth]{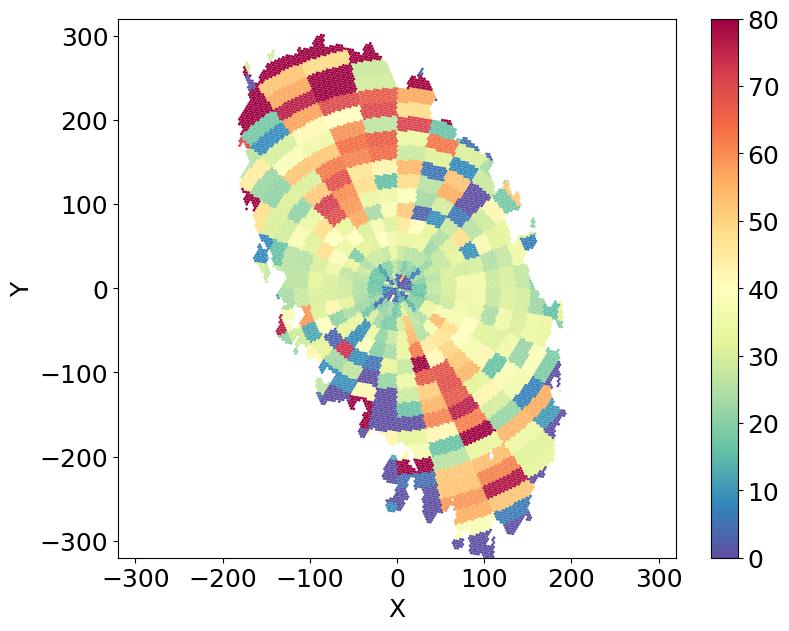}\\
\includegraphics[width=0.22\textwidth]{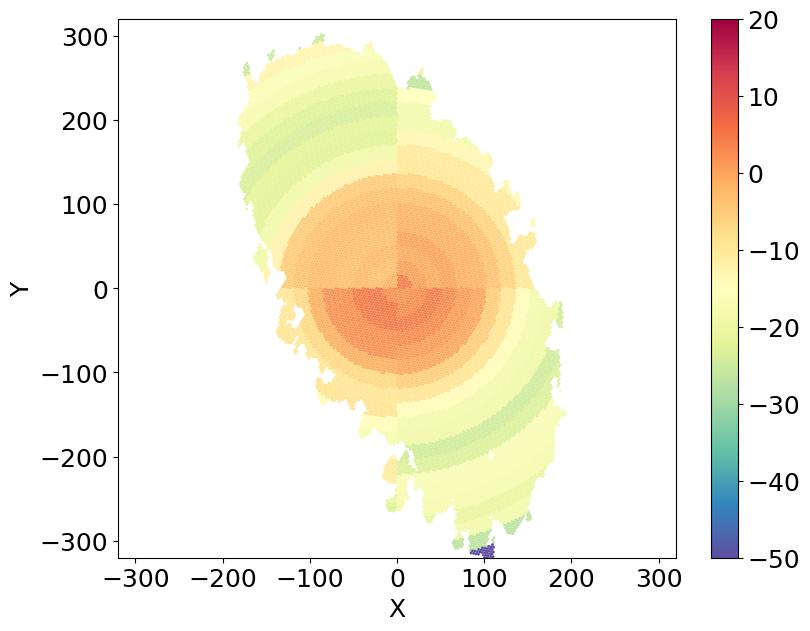}
\includegraphics[width=0.22\textwidth]{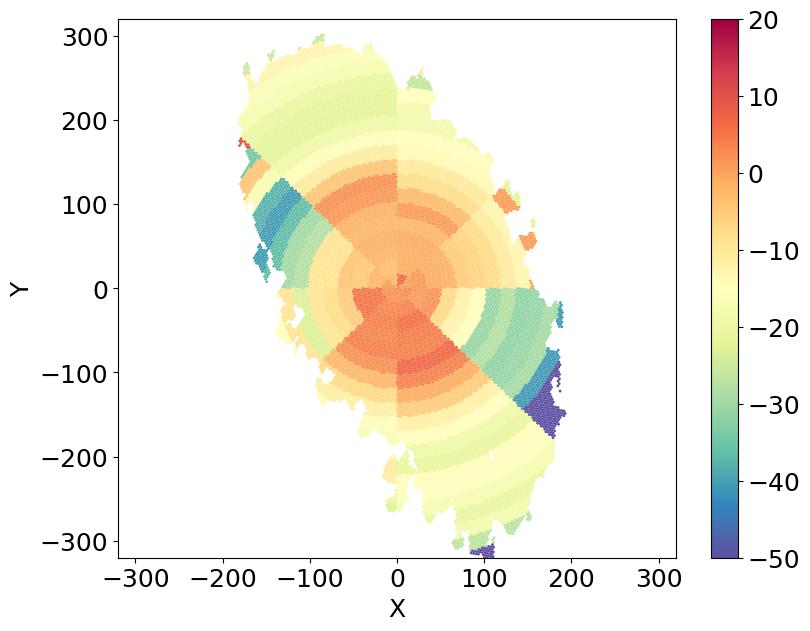}
\includegraphics[width=0.22\textwidth]{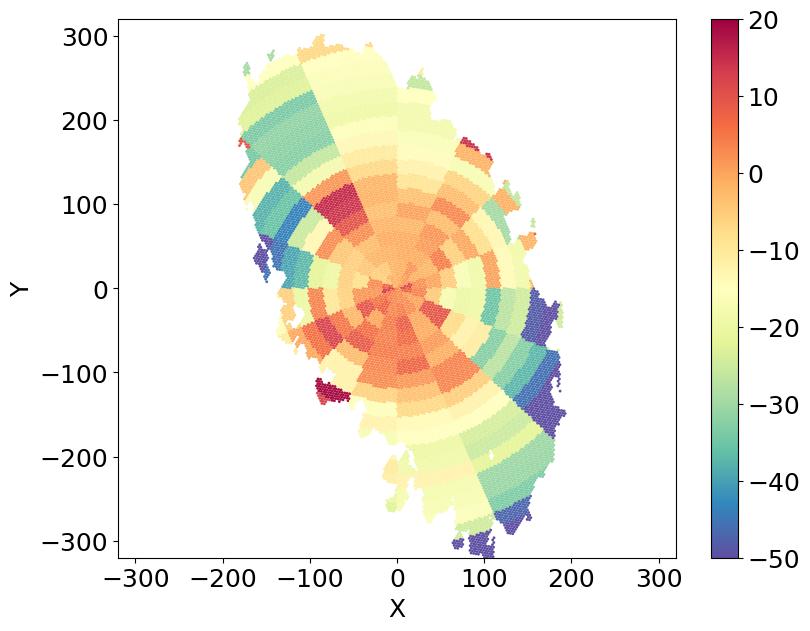}
\includegraphics[width=0.22\textwidth]{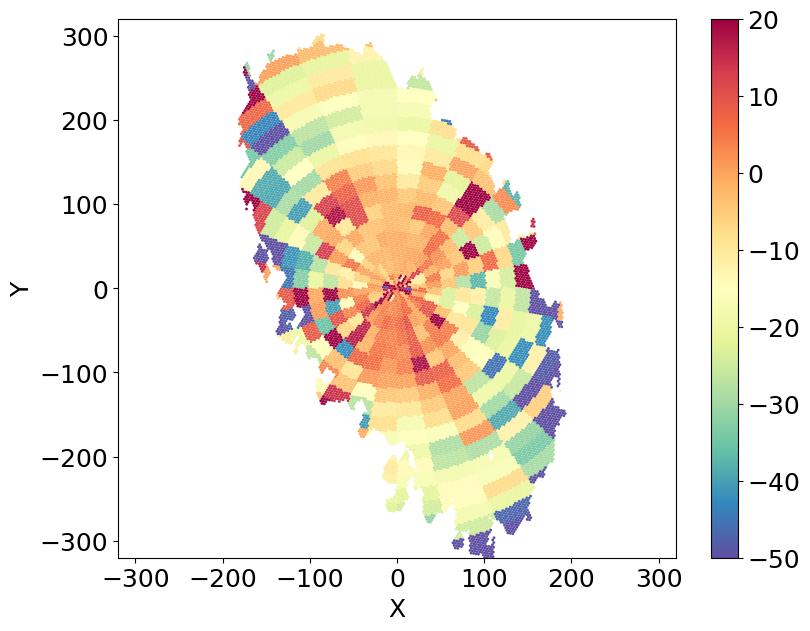}
\caption{ Results of the VRM model for $N_r=20$ concentric rings and $N_a=4, 8, 16, 32$ azimuthal sectors: respectively, upper panels show the transversal velocities while
bottom panels the radial ones.
{  The $X$ and $Y$ axes (the coordinates are in the plane of the galaxy) are in arc-seconds; the color code in km s$^{-1}$. }
} 
\label{vrm_res} 
\end{figure*}

As the method does not provide formal uncertainties for the velocity components in each cell, one way to control the effect of changing the resolution of the coarse-grained reconstructed map is to vary the number of cells. For instance, one can fix the number of concentric rings and vary the number of angular sectors (arcs). By doing so, it is possible to verify whether the velocity field in a given angular region (covered by at least one cell) yields consistent values of the velocity components regardless of the number of arcs used. Such consistency indicates that the maps of transverse and radial velocities have converged, and that the spatial distribution of anisotropies is robust with respect to the chosen angular resolution ({  see \citet{SylosLabini_etal_2023b} for more details)}.

Figure~\ref{vrm_res} shows the results obtained using \( N_r = 20 \) concentric rings while varying the number of azimuthal sectors with \( N_a = 4, 8, 16, 32 \), in order to explore the effects of angular resolution. These tests are specifically designed to verify that the velocity field reconstructed in a given angular region with a certain number of arcs \( N_a \) remains consistent with that obtained using a lower number of arcs.

The maps obtained with different values of \( N_a \) exhibit a consistent behavior: the anisotropies appear in the same regions of the galaxy, but become more sharply defined as the angular resolution increases. This behavior reflects the fact that, when \( N_a \) is small, the size of the individual cells is larger than the typical observational resolution \( \sigma_{res} \approx 10 '' \). {  This parameter characterizes} the minimum angular scale at which genuine features can be distinguished from spurious correlations introduced by instrumental smoothing. As long as the cell size exceeds \( \sigma_{res} \), such correlations do not significantly affect the signal across different cells.

To assess whether the VRM reliably reconstructs the velocity field on angular scales larger than \( \sigma_{res} \), we adopt the following strategy. We begin by fixing the number of rings and setting \( N_a = 1 \), corresponding to the monopole component. Then, we progressively double the number of arcs to \( N_a = 2, 4, 8, 16, 32 \). For each case, we verify that in a given angular region (covered by at least one cell at resolution \( N_a \)) the VRM yields consistent estimates of \( v_r \) and \( v_t \), independently of the number of arcs. Only under this condition can we conclude that the two-dimensional maps of the radial and tangential velocities converge, and that the recovered spatial distribution of anisotropies is robust.

It is important to emphasize that the convergence of the  moments is a necessary but not sufficient condition for assessing the reliability of the results. Specifically, convergence is required to determine whether the velocity field reconstructed with the VRM is trustworthy, in the sense that the noise inevitably affecting any measurement does not exceed the signal.

However, the effects introduced by geometric deformations,  such as a galactic warp, are fundamentally different, as they produce a real (i.e., non-noise-induced) signal. In such cases, the octopole moments may still converge, provided that the amplitude of the geometrically induced signal is larger than the noise. In other words, if convergence is not observed, then it is certain that measurement noise dominates over the signal. Conversely, if convergence is observed, this does not exclude the presence of a geometric deformation; it simply implies that the signal-to-noise ratio is high enough for meaningful interpretation.

As shown in Fig.~\ref{vrm_res}, the tangential velocity field is regular in the inner disk (\( R < 100'' \)), whereas it becomes more irregular and fluctuating in the outer regions. The radial velocity field exhibits a similar trend: it is negligible in the inner disk and displays more pronounced anisotropies in the outer parts.

\begin{figure*}
\centering
\includegraphics[width=0.44\textwidth]{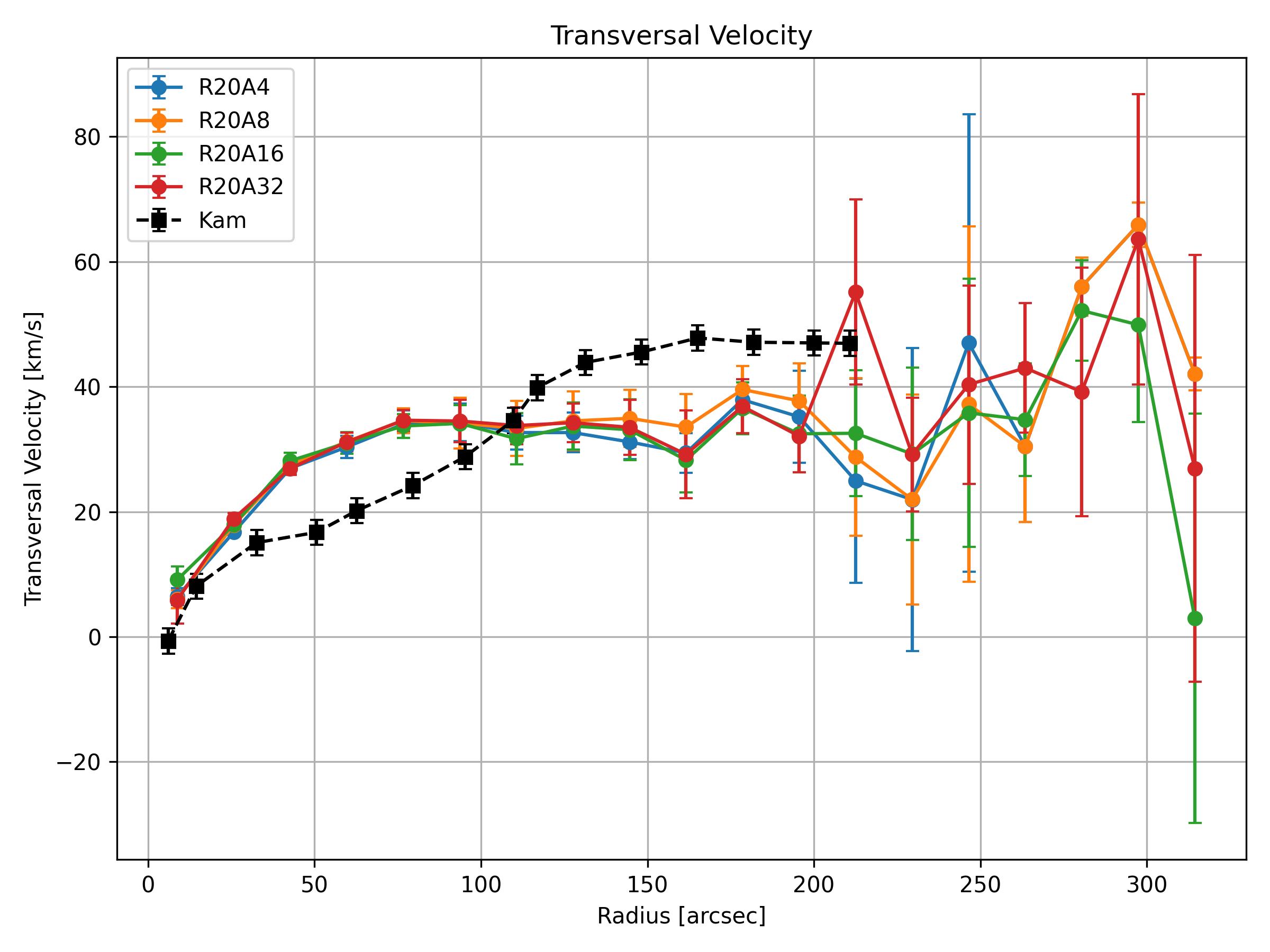}
\includegraphics[width=0.44\textwidth]{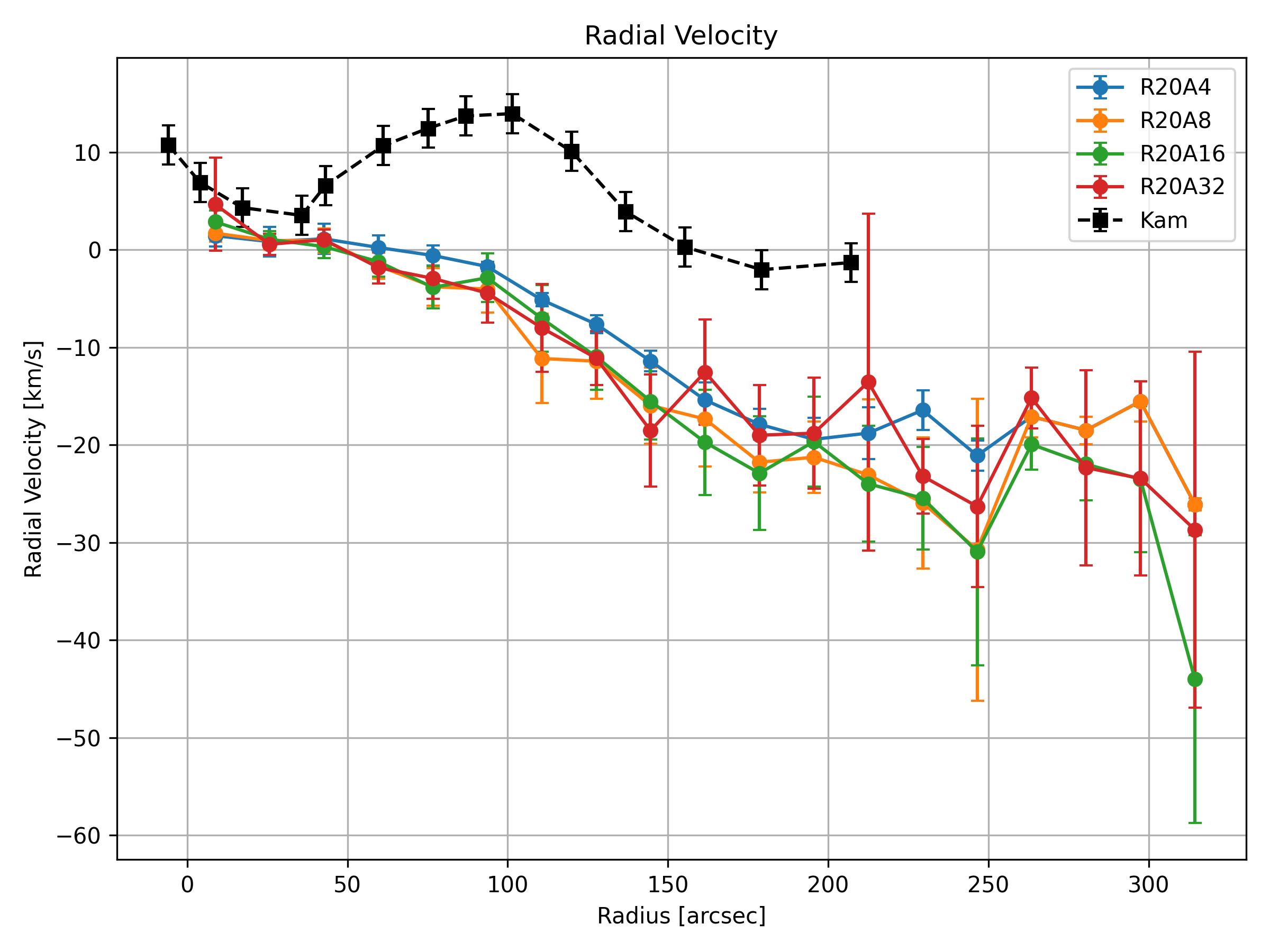}
\caption{Transversal (Left panel) and radial (Right panel) velocity profiles for  ESO 358-60: the number of rings is the same, $N_r=20$, whereas the number of arcs varies 
$N_a=4, 8, 16, 32$. Black dots (Kam) represents the results based on the TMR procedure (see text) {  obtained by \cite{Kamphuis_etal_2025}.}
 } 
\label{profiles_vrm} 
\end{figure*}

Figure~\ref{profiles_vrm} shows the tangential and radial velocity profiles, averaged over  rings, for ESO 358-60 at different angular resolutions, i.e., for different numbers of arcs. One can observe that, up to \( R \sim 150'' \), both profiles converge well, exhibiting consistent behavior across resolutions. At larger radii, however, fluctuations become more significant. In addition, for \( R > 150'' \), the amplitude of the radial motions reaches approximately half that of the tangential velocity, calling into question whether the disk can still be considered in rotational equilibrium. For this reason we will limit the to $R=150''$ the analysis for the estimation of the galaxy's mass.

Finally, it should be emphasized that the differences observed in both the tangential and radial velocity profiles arise from the distinct assumptions about disk geometry adopted by the two models. The fact that radial velocities become more prominent in the outermost regions of the galaxy is consistent with previous findings in several other systems \citep{SylosLabini_etal_2024, SylosLabini_etal_2025a}, and supports the idea that these outer regions are not fully relaxed. This may be attributed either to the increasing influence of tidal forces or to the dynamical history of the system: in particular, during the collapse phase, {  a fraction of the galaxy's mass} may have gained sufficient kinetic energy to move on highly eccentric orbits. We will return to this point {  below}. In contrast, we stress that  the peak in \( v_r \) observed in the profile derived using the TRM-based procedure appears rather anomalous and may be an artifact arising from the geometric assumptions inherent to that method.

%%%%%%%%%%%%%%%%%%%%%%%%%%%%%%%%%%%%%%%%%%%%%%%%%%%%%%%%%%%%%%%%%

\subsection{Cross correlation} 

Once the tangential $v_t(R,\theta)$ and radial $v_r(R,\theta)$ velocity fields have been reconstructed, we compute their cross-correlations with each other, as well as with the velocity dispersion field $\sigma(R,\theta)$ and the surface brightness field $\Sigma(R,\theta)$. In addition, we evaluate the cross-correlation between $\sigma(R,\theta)$ and $\Sigma(R,\theta)$. {  These correlation maps provide insight into the structure and dynamical nature of the galaxy's velocity field, as discussed in the following sections.}

To quantify the cross-correlation between two generic fields \( f(R, \theta) \) and \( g(R, \theta) \), we consider the Pearson correlation coefficient (i.e., the correlation coefficient between the rank variables) defined as (see, e.g., \citealt{Numerical_Recepies,SylosLabini_etal_2025b}):
\begin{equation}
\label{crosscorrcoeff}
r_{fg} = \frac{1}{N_{\mathrm{cells}} - 1} \sum_{i=1}^{N_{\mathrm{cells}}} r_{fg}^i(R, \theta),
\end{equation}
where
\begin{equation}
\label{crosscorrcoeff_2}
r_{fg}^i(R, \theta) = \left( \frac{f^i(R, \theta) - \overline{f}}{s_f} \right) \left( \frac{g^i(R, \theta) - \overline{g}}{s_g} \right),
\end{equation}
and \( f^i(R, \theta) \) denotes the value of the field \( f \) in the \( i^{\mathrm{th}} \) cell, with \( i = 1, \ldots, N_{\mathrm{cells}} \).

The mean and standard deviation of the field \( f \) are defined as
\begin{align}
\label{crosscorrcoeff_3}
\overline{f} &= \frac{1}{N_{\mathrm{cells}}} \sum_{i=1}^{N_{\mathrm{cells}}} f^i(R, \theta), \\[4pt]
s_f^2 &= \frac{1}{N_{\mathrm{cells}} - 1} \sum_{i=1}^{N_{\mathrm{cells}}} \left( f^i(R, \theta) - \overline{f} \right)^2,
\end{align}
and analogously for \( \overline{g} \) and \( s_g \).

We also employ Spearman correlation coefficients  which are defined as the Pearson correlation coefficients between the rank variables \citep{Numerical_Recepies,SylosLabini_etal_2025b}.

\begin{figure} 
\includegraphics[width=0.42\textwidth]{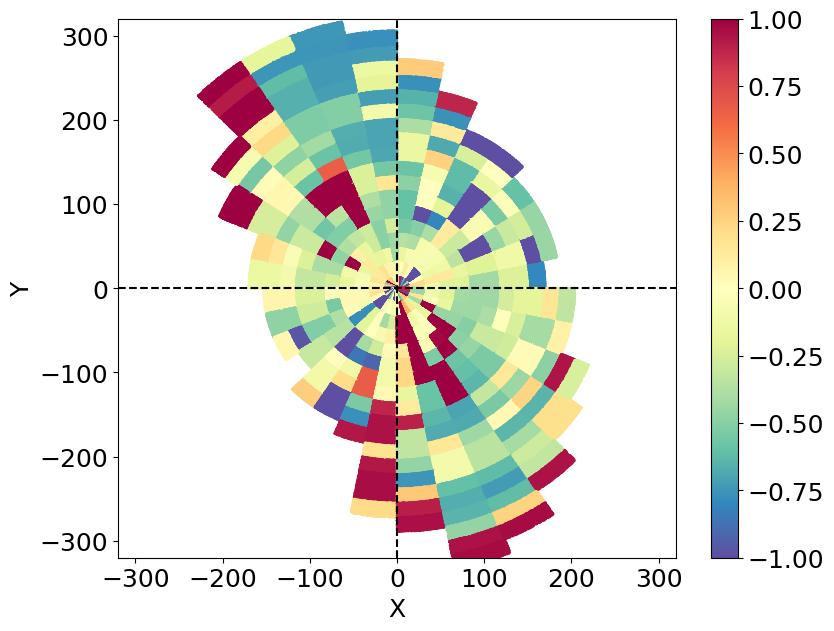}
\includegraphics[width=0.42\textwidth]{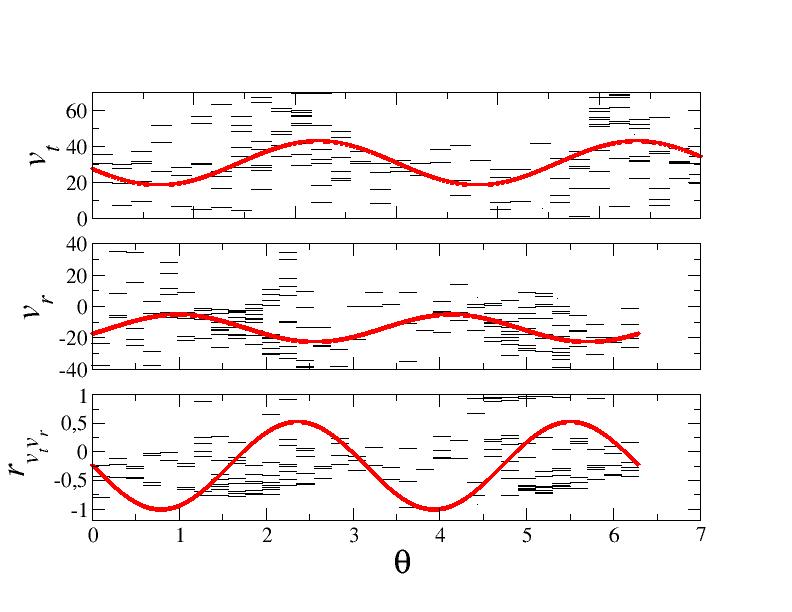}
\caption{
{Upper panel:} Rank velocity correlation map for ESO 358-60, computed with \( N_r = 20 \) rings and \( N_a = 32 \) angular sectors.  
{  The $X$ and $Y$ axes (the coordinates are in the plane of the galaxy) are in arc-seconds.  }
{Bottom panel:} Tangential velocity component, radial velocity component, and rank velocity correlation coefficient as functions of the polar angle \( \theta \), measured in the outer region of the ESO 358-60 disk (\( R > 150'' \)).
} 
\label{rank_vel_map} 
\end{figure}

The map of the velocity rank correlation is shown in the upper panel of Fig.~\ref{rank_vel_map}; one can observe that no clear dipolar correlation is present. This observation is confirmed by the analysis presented in the bottom panel of Fig.~\ref{rank_vel_map}. As discussed below (see also \citealt{SylosLabini_etal_2025b}), a dipolar correlation is expected if the warp is a genuine feature of the disk, rather than an artifact introduced by the TRM method.

\begin{figure} 
\includegraphics[width=0.42\textwidth]{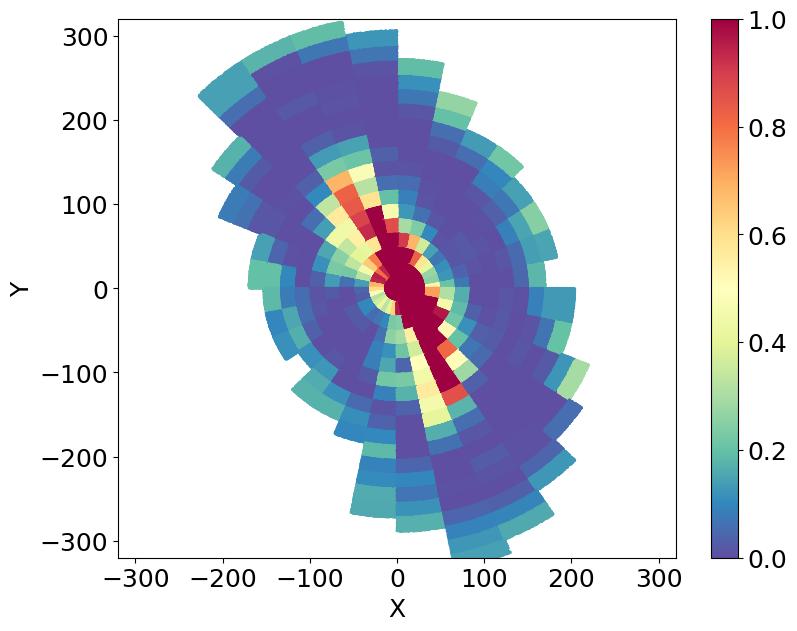}
\includegraphics[width=0.42\textwidth]{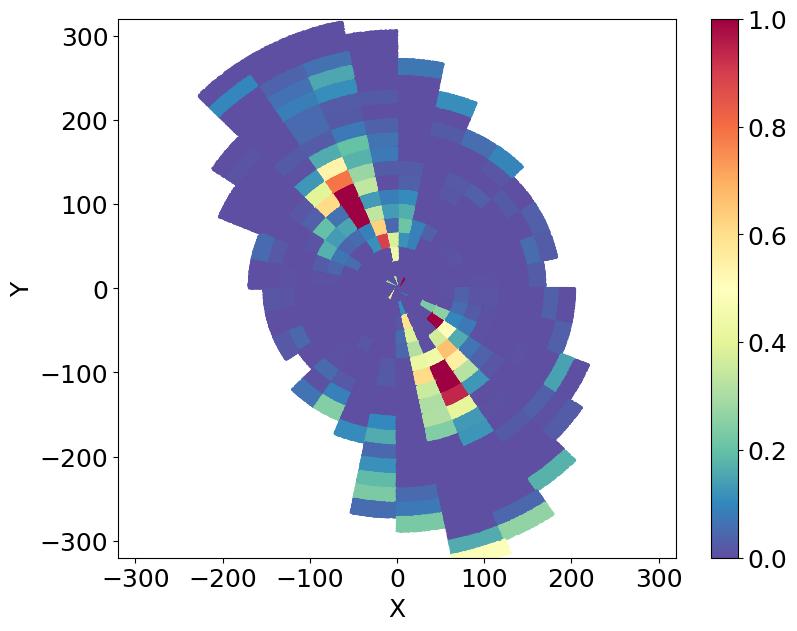}
\includegraphics[width=0.42\textwidth]{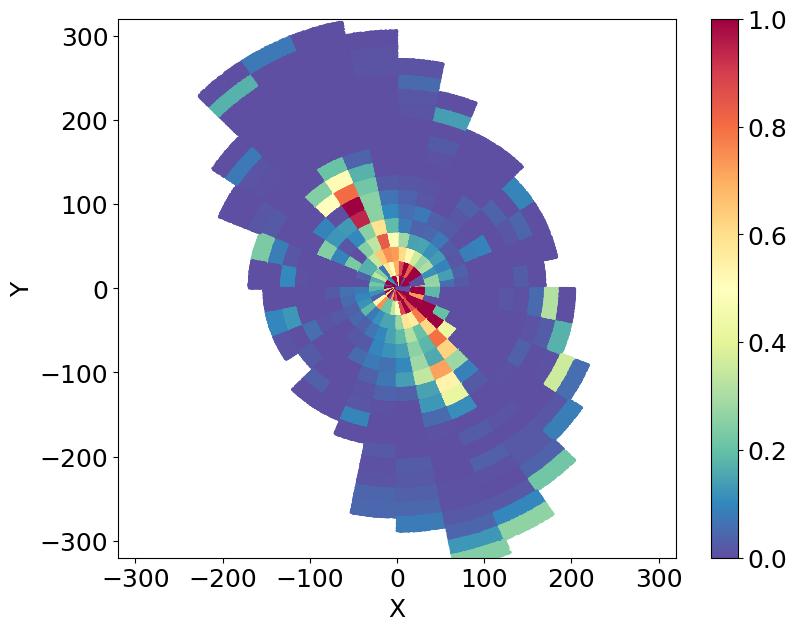}
\caption{
Upper panel: correlation map between surface brightness and velocity dispersion. 
Middle panel: correlation map between surface brightness and radial velocity. 
Bottom panel: correlation map between surface brightness and transversal velocity dispersion. 
All correlation maps have been computed using \(N_r = 20\) radial rings and \(N_a = 32\) angular sectors.
{  The $X$ and $Y$ axes (the coordinates are in the plane of the galaxy) are in arc-seconds.  }
} 
\label{SB-disp_map} 
\end{figure}

Finally, Fig.~\ref{SB-disp_map} shows the surface brightness-velocity dispersion correlation map. The central over-density in surface brightness is highly correlated with the velocity dispersion map, providing further evidence that this structure corresponds to a bar rather than to an almost edge-on disk. Indeed, as rotation takes place in a 2D disc, the projected component of the rotation velocity decreases in amplitude with decreasing inclination, whereas the contribution due to random motions, which take place in 3D, remains the same by changing the inclination angle \citep{SylosLabini_etal_2024}.
{ 
Fig.~\ref{SB-disp_map} presents also the surface-brightness–radial velocity and surface-brightness–transversal velocity correlation maps. All three maps consistently show an enhanced correlation along the bar for $R < 100''$. This strengthened correlation between the intensity and the velocity components supports the presence of a bar, as such a structure is dynamically expected to leave a clear imprint on the kinematic field.
} 

\subsection{Test on warp} 
 
 In \cite{SylosLabini_etal_2025b}, we implemented a procedure to assess the validity of the warp identified by the TRM in a galactic disk. The first step consists of applying the TRM method to measure the circular velocity, the inclination angle, and the P.A. Using these radial profiles, we construct a toy disk galaxy model that reproduces the observed geometric deformation inferred from the TRM and the observed rotation curve, while assuming zero radial velocity.
\begin{figure} 
\includegraphics[width=0.42\textwidth]{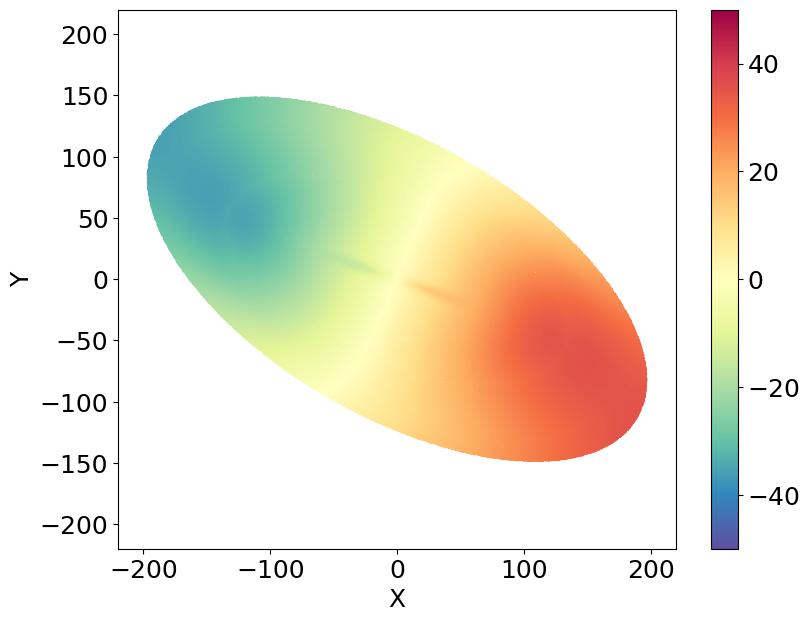}
\caption{The 2D line-of-sight velocity field of the toy model used as a reference.
{  The $X$ and $Y$ axes (the coordinates are in the plane of the sky) are in arc-seconds; the color code in km s$^{-1}$. }
} 
\label{vlos_toy} 
\end{figure}
Given this model, we generate a synthetic LOS velocity 2D map, as it would appear to a distant observer, adopting the global inclination angle measured from observations (see Fig.\ref{vlos_toy}). We then apply the VRM method to this synthetic map to reconstruct the transverse and radial velocity components, as well as their rank correlation coefficient. By construction, both velocity components and the correlation coefficient exhibit a clear dipolar modulation in the VRM-reconstructed maps, as this is a characteristic feature of a warp encoded in the toy model.
Then, if the rank correlation coefficient derived from the real galaxy closely resembles the dipolar modulation observed in the toy model, then the disk is likely warped, thereby validating the TRM reconstruction as a faithful representation of the system's properties.

In summary, the concept behind this procedure is to use, for {  a given} galaxy, its corresponding toy model (constructed from the velocity profiles and geometric deformation obtained through the TRM) as a null hypothesis test. This approach enables us to isolate features in the velocity maps that cannot be attributed solely to a warp, and which therefore correspond to extrinsic velocity perturbations.

\begin{figure} 
\includegraphics[width=0.42\textwidth]{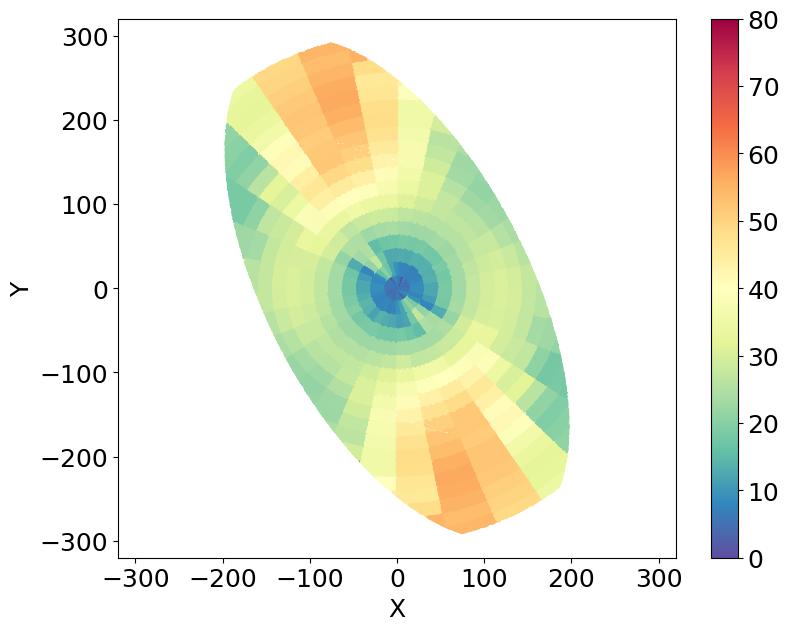}
\includegraphics[width=0.42\textwidth]{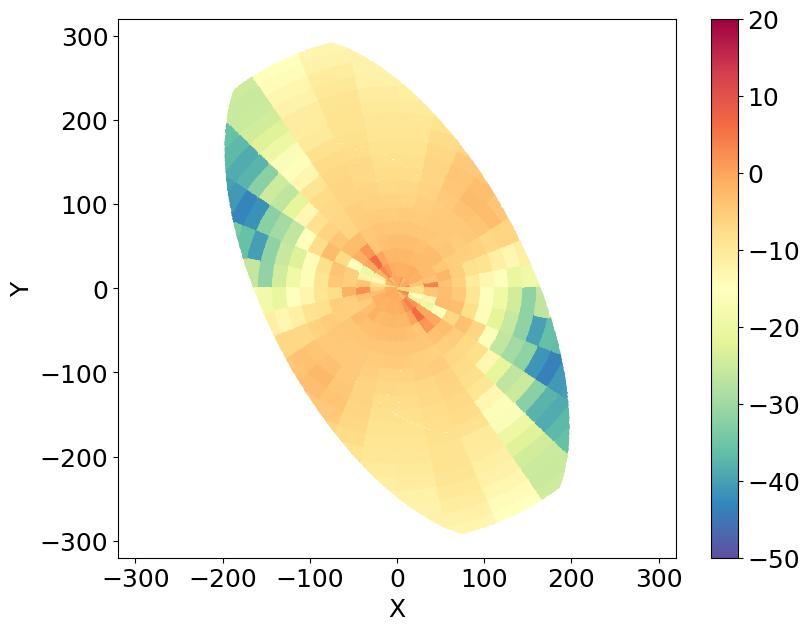}
\caption{ As Fig.\ref{vrm_res} but for the toy model  computed with \( N_r = 20 \) rings and \( N_a = 32 \) angular sectors.
} 
\label{vrm_mod_res} 
\end{figure}

The VRM-reconstructed tangential and radial velocity fields, computed with \( N_r = 20 \) rings and \( N_a = 32 \) angular sectors, are shown in Fig.~\ref{vrm_mod_res}. One can observe that both components exhibit a dipolar angular modulation, although with different phases. This modulation was not found in the observed data for ESO 358-60, as illustrated in Fig.~\ref{vrm_res}.

\begin{figure} 
\includegraphics[width=0.422\textwidth]{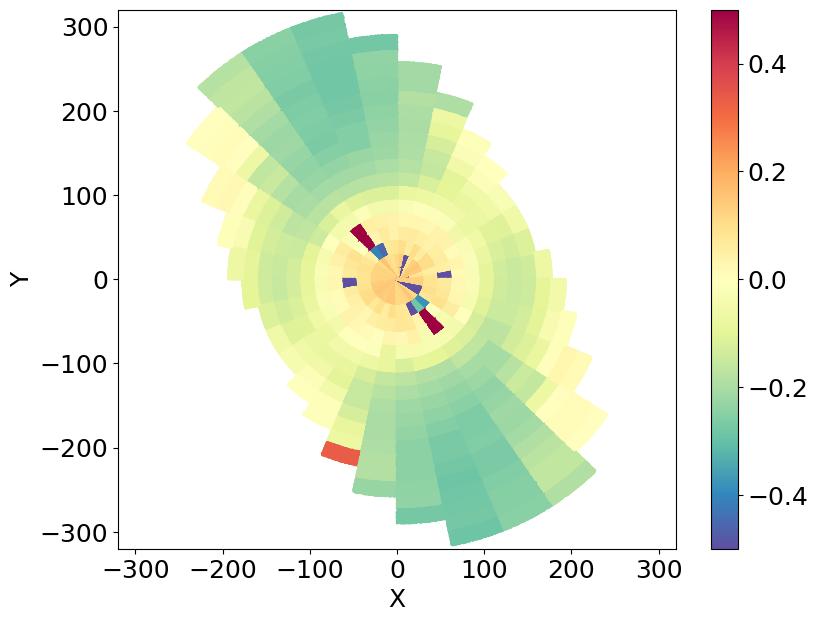}
\includegraphics[width=0.422\textwidth]{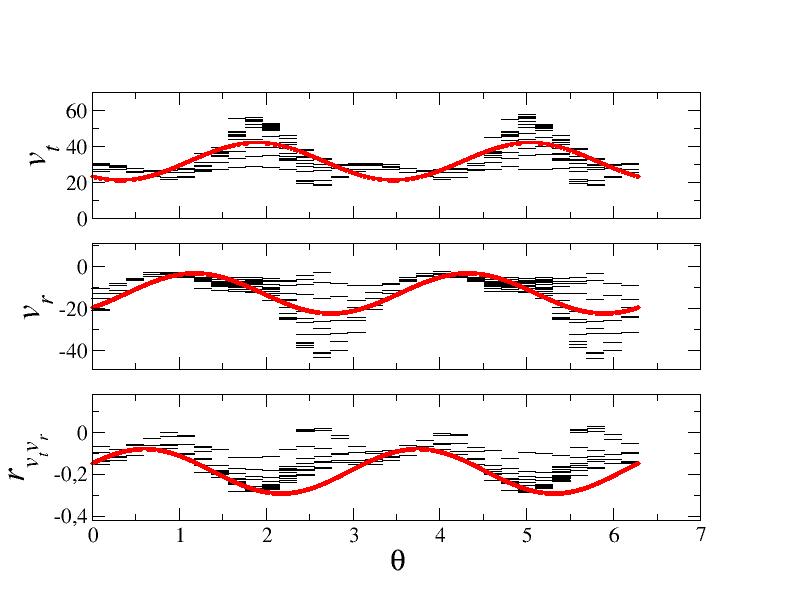}
\caption{As Fig.\ref{rank_vel_map}  but for the toy model; note, however, that the scale of the y-axis is different in all the panels.} 
\label{rank_vel_map_toy}  
\end{figure}

Figure~\ref{rank_vel_map_toy} presents the analysis for the toy model: the upper panel shows the rank velocity correlation map, and the lower panel displays the tangential velocity component, radial velocity component, and the rank correlation coefficient as functions of the polar angle \( \theta \), measured in the outer region of the disk (\( R > 150'' \)).

Two key observations emerge:
\begin{itemize}
    \item Both velocity components show a dipolar modulation with different phases.
    \item The rank velocity correlation coefficient also exhibits a dipolar pattern, although with small amplitude due to the phase difference between the two velocity components.
\end{itemize}
These features are absent in the real data for ESO 358-60 (see Fig.~\ref{rank_vel_map}). For this reason, we conclude that the warp inferred by the TRM is likely an artifact of the method rather than a genuine structural feature of the galaxy.

%%%%%%%%%%%%%%%%%%%%%%%%%%%%%%%%%%%%%%%%%%%%%%%%%%%%%%%%%%%%

\subsection{Discussion} 

In our analysis, the VRM adopts the simplifying assumption that the galactic disk is flat, i.e. without a warp, so that both the inclination and position angles remain constant with radius. This choice is motivated by tests with toy galaxy models, which indicate that, in the case of ESO 358-60, the presence of a warp is not supported. In particular, a warp would manifest as a correlation with the rank velocity correlation coefficient, which is not observed in the data. The VRM interpretation is thus based on, and consistent with, this stringent observational test. At the same time, this test leads us to conclude that the TRM analysis of  \citet{Kamphuis_etal_2025}   does not provide a consistent description of the kinematic data.

The optical data (see the right panel of Fig.~\ref{fig:M0M2}) reveal only the inner stellar structure. While \citet{Kamphuis_etal_2025} interpret this feature as an inner disk with inclination $i = 80^\circ$, in our analysis it is more naturally identified as a bar-like structure associated with the central \HI{} overdensity (see the left panel of Fig.~\ref{fig:M0M2}). In either interpretation, the optical image provides no evidence for the presence of two distinct disk components---namely an inner and an outer disk with inclination angles differing by $\sim 20^\circ$.

Moreover, the optical image does not convey information beyond that already contained in the \HI{} intensity map, as the optical overdensity coincides spatially with the region of highest stellar concentration. The linear extent of this feature is significantly smaller than $100''$, whereas the possible warp inferred by \citet{Kamphuis_etal_2025} from the TRM analysis (see their Fig.~5b) appears only at radii larger than $100''$, well beyond the extent of the stellar component detected in the optical.

Finally, the \HI{} disk at inclination $i = 60^\circ$---present both in the TRM model of \citet{Kamphuis_etal_2025} for $R > 150''$ and in our model across the full disk---is not detected in the optical image. This is expected, as this component is dominated by neutral hydrogen and contains only a minor stellar contribution. The stellar component is instead confined to the region $R < 100''$, where the \HI{} overdensity corresponds to the elongated optical feature.

We also note that the stellar structure visible in Fig.~\ref{fig:M0M2}, together with the associated \HI{} overdensity, is not perfectly linear at its edges. However, galactic bars are rarely strictly linear; modest deviations from straightness, particularly in their outer regions, can naturally arise due to substructure or local asymmetries. In this context, the observed morphology is fully consistent with a bar-like interpretation.

More generally, barred galaxies are common among disk systems. Results from the Galaxy Zoo project \citep{Masters_etal_2011} indicate that approximately 30\% of galaxies in their sample host a bar, in good agreement with earlier visually classified surveys. This broader empirical context further supports our interpretation that the observed structure in ESO~358--60 is plausibly explained by the presence of a bar.

 %%%%%%%%%%%%%%%%%%%%%%%%%%%%%%%%%%%%%%%%%%%%%%%%%%%%%%%
%%%%%%%%%%%%%%%%%%%%%%%%%%%%%%%%%%%%%%%%%%%%%%%%%%%%%%%
%%%%%%%%%%%%%%%%%%%%%%%%%%%%%%%%%%%%%%%%%%%%%%%%%%%%%%%

\section{Dynamical mass model} 
\label{sec:mass_est} 

{ 
Assuming a distance of \( D = 9.4 \pm 2.5 \,\mathrm{Mpc} \), the stellar mass of the galaxy, derived from the 3.6~\(\mu\)m flux, is estimated as  
\[
M_{\ast} = (3 \pm 2) \times 10^{7}\, M_\odot \, .
\]  
The neutral hydrogen mass is  
\[
M_{\text{H\sc i}} = (3 \pm 1) \times 10^{8}\, M_\odot \, .
\]  
Including the contribution of helium, obtained by applying a correction factor of 1.3, the total gaseous mass becomes  
\[
M_{\text{gas}} = (3.9 \pm 1.3) \times 10^{8}\, M_\odot \, .
\]  
} 

 The significant uncertainties on these mass estimates arise primarily from the uncertainty in the distance \citep{Kamphuis_etal_2025}. Following \cite{SylosLabini_etal_2024, SylosLabini_etal_2025a} we estimate, under two different assumptions, the total mass which is the sum of the dark matter mass and the baryonic mass
\[
M_{\text{bar}} = M_{\text{g}} + M_{\text{s}} =  (3.3 \pm 1) \times 10^8  M_\odot \;.
\] 
As a first model, we consider the standard NFW  halo profile \citep{Navarro_etal_1997}, in which dark matter is distributed in a spherical halo characterized by two parameters: the central density \( \rho_0 \) and the scale radius \( r_s \). The density profile is given by
\begin{equation}
\label{rho_nfw} 
\rho(r) = \frac{\rho_0}{\left( \frac{r}{r_s} \right) \left( 1 + \frac{r}{r_s} \right)^2},
\end{equation}
where \( r \) denotes the spherical radius.

In this framework, the total rotational velocity is expressed as
\begin{equation}
\label{vc_nfw} 
v_{\text{c}}^2(R) = v_{\text{s}}^2(R) + v_{\text{g}}^2(R) + v_{\text{h}}^2(R),
\end{equation}
where \( v_{\text{s}}(R) \) and \( v_{\text{g}}(R) \) are the contributions from the stellar and gas components, respectively, and \( v_{\text{h}}(R) \) is the contribution from the dark matter halo. Such circular velocities  are defined as the equilibrium rotational velocities that each component would induce on a test particle in the plane of the galaxy, assuming it were isolated. The circular velocity of the halo, \( v_{\text{h}}(R) \), can be computed analytically from the NFW density profile (Eq.\ref{rho_nfw}).

In contrast, the circular velocities of the disk components, \( v_{\text{s}}(R) \) and \( v_{\text{g}}(R) \), must be determined numerically. More specifically, we generate a numerical realization of a thin-disk model using the observed surface brightness profile of the gas component and its corresponding total mass. From this, we compute the resulting centripetal acceleration numerically. Hereafter, we neglect the contribution of the stellar component, \( v_{\text{s}}(R) \), as its mass is approximately one-tenth that of the gaseous component and therefore has a negligible dynamical effect.

\begin{figure} 
\includegraphics[width=0.42\textwidth]{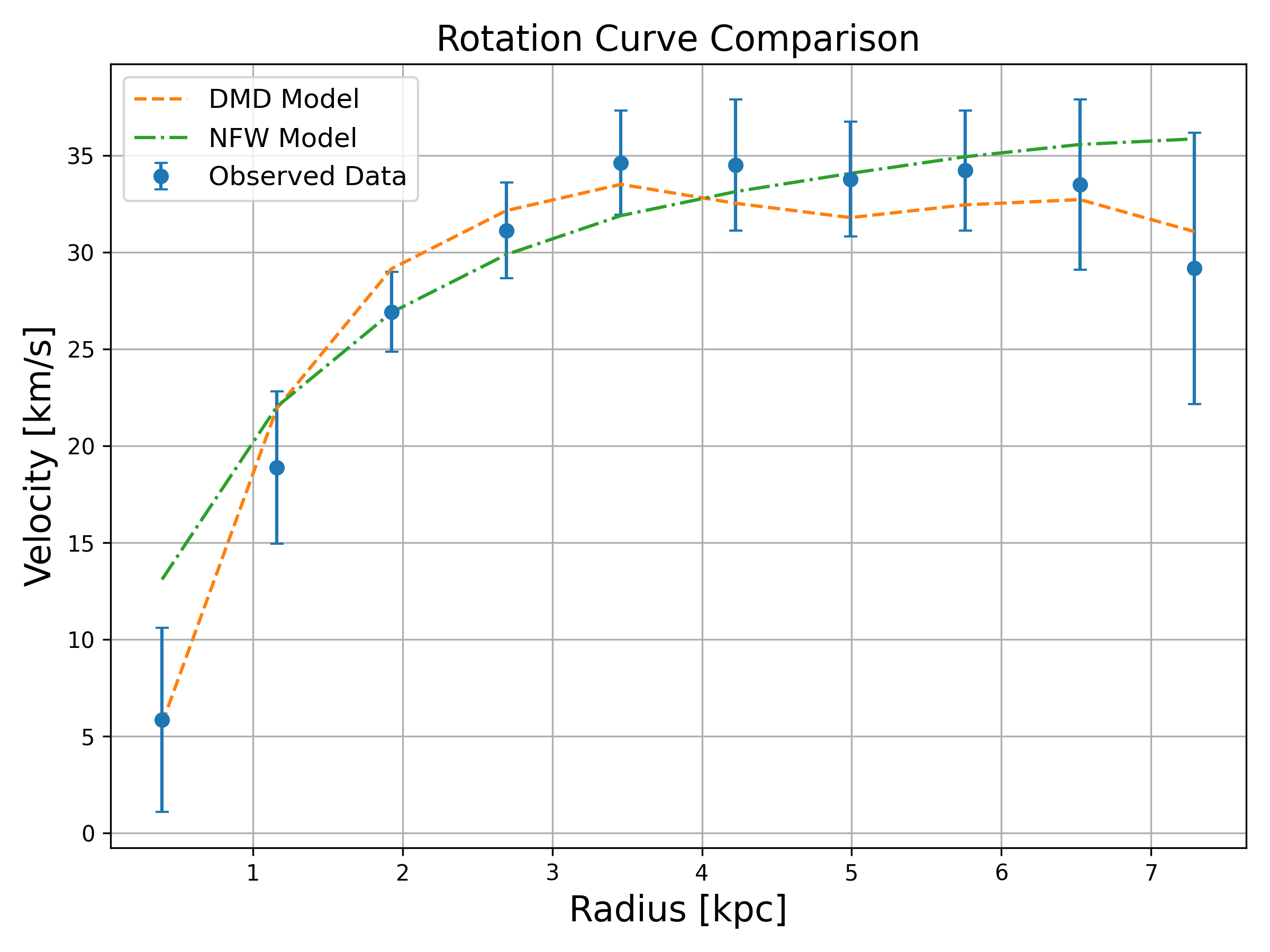}
\caption{Best fit with a NFW and DMD model of ESO 358-60 rotation curve obtained with the VRM. } 
\label{mass_fit}  
\end{figure}
The total mass of the baryonic disk plus the halo is given by
\begin{equation}
\label{M_nfw}
M_{\mathrm{NFW}} = M_{\mathrm{bar}} + M_{200},
\end{equation}
where \( M_{\mathrm{bar}} \) is the baryonic mass, and \( M_{200} \) is the mass of the dark matter halo enclosed within a sphere of radius \( r_{200} \), defined as the virial radius \citep{Navarro_etal_1997}.

We then consider the DMD model, which assumes that dark matter is confined to the galactic disk \citep{Hessman+Ziebart_2011, SylosLabini_etal_2023b, SylosLabini_etal_2024, SylosLabini_etal_2025a}. In this framework, the total rotational velocity is given by
\begin{equation}
\label{vc_dmd}
v_{\text{c}}^2(R) = \gamma_{\text{s}} v_{\text{s}}^2(R) + \gamma_{\text{g}} v_{\text{g}}^2(R),
\end{equation}
where \( \gamma_{\text{s}} \) and \( \gamma_{\text{g}} \) are, in general, free parameters representing the relative contribution of dark matter associated with the stellar and gas components, respectively.
In the present case, given that the stellar mass component is negligible, we set \( \gamma_{\text{s}} = 0 \), leaving \( \gamma_{\text{g}} \) as the sole free parameter of the model.

As in the case of the halo model in Eq.~\ref{vc_nfw}, each velocity component in Eq.~\ref{vc_dmd} represents the rotational velocity induced by that component on a test particle in the plane of the galaxy, assuming the component is isolated. These velocities were computed numerically as described above.

In this framework, the total mass of the baryonic and dark matter components is given by
\begin{equation}
\label{M_dmd}
M_{\mathrm{DMD}} = \gamma_{\text{s}} M_{\text{s}} + \gamma_{\text{g}} M_{\text{g}} \approx \gamma_{\text{g}} M_{\text{g}} \;,
\end{equation}
where the approximation holds because the stellar mass \( M_{\text{s}} \) is negligible compared to the gas mass \( M_{\text{g}} \).

The best-fit parameters for the halo model are
\begin{align}
r_s &= 6.3 \;\; \text{kpc}, \\
\rho_0 &= 1.84 \times 10^{-25} \;\; \text{g cm}^{-3}. \nonumber
\end{align}
The reduced chi-squared value, with two degrees of freedom, is \( \chi^2_\nu = 0.70 \). The corresponding virial mass is estimated to be
\[
M_{\mathrm{vir}} \approx 1.1 \times 10^{10} \;\; M_\odot.
\]
For the DMD model, which has a single free parameter, the best-fit yields \( \chi^2_\nu = 0.38 \). We find
\[
{  \gamma_{\text{g}} = 6.5,}
\]
corresponding to a total mass of
\[
M_{\mathrm{DMD}} = (2.5 \pm 0.8) \times 10^9 \;\; M_\odot.
\]
Note that the 30\% uncertainty on the total mass estimate arises from the uncertainty in the baryonic mass determination.

In conclusion, the DMD model provides a statistically better fit than the NFW model, as indicated by a reduced chi-squared value that is approximately half. In addition, the total mass inferred from the DMD model is about four times smaller than the virial mass obtained from the NFW fit.

\section{On the physical origin of dark matter disks} 
\label{sect:orgin}

{ 

Let us now discuss in greater detail the physical motivations underlying the choice of the two mass models considered above.

\subsection{Formation of disks in CDM models} 

Within the standard $\Lambda$CDM paradigm, non-linear structure formation proceeds hierarchically through a bottom-up process, in which small-scale density fluctuations merge and accrete to form progressively larger systems. The resulting nonlinear structures, the dark matter halos, have nearly spherical configurations and quasi-isotropic velocity dispersions. Rotationally supported disks are not a direct outcome of gravitational dynamics alone, but instead emerge from dissipative, non-gravitational processes acting on the baryonic component, as originally envisaged by \cite{ELS_1962}. Gas can undergo shocks, radiate energy, and collapse while approximately conserving angular momentum, thereby forming a thin, rotationally supported disk embedded within the much larger gravitational potential of the dark matter halo. 

As a consequence of these distinct formation pathways, baryonic and dark matter components are expected to display markedly different spatial and kinematic properties. Baryons, subject to dissipation, settle into a flattened rotating disk, whereas dark matter---being collisionless and dissipationless---forms a quasi-spherical halo characterized by an approximately isotropic velocity dispersion and a total mass typically exceeding that of the baryonic component by more than an order of magnitude. 

In the standard picture, the dark matter halo provides the gravitational potential that stabilizes the disk. When the disk mass becomes a significant fraction of the halo mass, however, this stabilizing role is no longer guaranteed, and the dynamical stability of the disk may be compromised. This raises a fundamental question regarding the stability of massive, disk-dominated systems.

In $\Lambda$CDM cosmological simulations, such non-linear halo structures are commonly described by the Navarro--Frenk--White (NFW) density profile \citep{Navarro_etal_1997}, which predicts a cuspy inner density distribution, $\rho(r) \propto r^{-1}$ 
\citep{Moore_1994, Navarro_etal_1996, Navarro_etal_1997, Moore_etal_1999, Navarro_etal_2004, Navarro_etal_2010, Power_etal_2003, Diemand_etal_2008, Stadel_etal_2009, Ishiyama_etal_2013}. Observationally, however, many dwarf galaxies exhibit rotation curves that rise approximately linearly toward the center, $v_c \propto r$, implying a constant-density core, $\rho(r) \approx \mathrm{const.}$ 
\citep{Moore_1994, deBlok_etal_1996, deBlok_etal_2001, deBlok+Bosma_2002, Oh_etal_2008, Oh_etal_2011}. This discrepancy between the cuspy inner profiles predicted by $\Lambda$CDM simulations and the cored profiles inferred from observations---the so-called cusp--core problem---represents a central aspect of the small-scale challenges faced by the $\Lambda$CDM model \citep[e.g.][]{Kroupa_2012, Pawlowski+Kroupa_2013, Dabringhausen+Kroupa_2103, Sales_etal_2022}.

In principle, the distinction between cusped and cored density profiles is of fundamental importance, as it may signal different underlying dynamical formation mechanisms. In practice, however, this distinction becomes less clear once additional physical processes are taken into account. For instance, \cite{Read_etal_2016} showed that stellar feedback can heat dark matter, producing a modified coreNFW profile that is consistent with the observed rotation curves of dwarf galaxies. Similarly, \cite{Lazar_etal_2020} demonstrated that the inclusion of realistic galaxy formation physics can yield dark matter halos that are either cored or cuspy. Modeling galaxy evolution therefore requires incorporating a complex interplay of gravity, hydrodynamics, radiative cooling and heating, and stellar feedback \citep{Sales_etal_2022}.

\subsection{Quasi-stationary disks from monolithic collapse}

An alternative pathway for non-linear structure formation is offered by top-down, monolithic collapse scenarios, in which bound systems emerge through rapid and violent gravitational dynamics \citep{Lynden_Bell_1967, Joyce_etal_2009}. In initially far-from-equilibrium systems, violent relaxation drives the evolution toward a quasi-stationary state (QSS) within a few dynamical timescales. If the initial conditions are anisotropic, departures from spherical symmetry are naturally amplified during collapse \citep{Lin_Mestel_Shu_1965}, allowing flat, disk-like structures to form purely via dissipationless gravitational interactions.

When the collapsing system possesses non-zero angular momentum, the resulting QSS can become rotationally supported and may develop long-lived spiral features \citep{Benhaiem+Joyce+SylosLabini_2017, Benhaiem+SylosLabini+Joyce_2019}. The inclusion of dissipative baryonic physics further enhances this process, leading to thinner, more coherent disks \citep{SylosLabini_etal_2020}.

The quasi-stationary nature of these disk-like configurations arises directly from the dynamics of gravitational collapse. In contrast to standard analyses of disk stability---which typically assume an initially equilibrated, rotationally supported configuration and assess its susceptibility to perturbations through linear theory or $N$-body simulations (see, e.g., \citealt{Toomre_1964, Merritt+Sellwood_1994, Athanassoula_2003, Revaz+Pfenniger_2004, Revaz_etal_2009, Peschken+Lokas+Athanassoula_2017})---stability in this context is neither externally imposed nor dependent on fine-tuned initial conditions. Rather, it emerges self-consistently from the non-linear gravitational evolution that drives the system toward a quasi-stationary state (QSS).

During monolithic collapse, violent relaxation produces a broad energy distribution that leads to two key outcomes: (i) the emergence of strongly turbulent velocity fields and (ii) the ejection of a significant fraction of the system's mass. As a result, the mass of the resulting disk is typically only a subset of the initial overdensity \citep{Benhaiem+Joyce+SylosLabini_2017}. The pronounced velocity gradients that characterize the QSS further enhance its resilience to perturbations, contributing to its long-lived stability.

This relaxation mechanism departs fundamentally from the classical framework of \citet{Lynden_Bell_1967}, where mass and energy are strictly conserved and the system evolves toward a maximum-entropy configuration. In the case of monolithic collapse, by contrast, these conservation laws are violated within the bound subsystem \citep{Joyce_etal_2009}, and the final QSS retains only a dynamically selected fraction of the original phase-space distribution. Since gravity lacks an intrinsic scale, this mechanism operates uniformly across a broad range of mass scales.

Such top-down collapse scenarios may  be realized in a cosmological context through suitable modifications of the small-scale behavior of the primordial power spectrum \citep{SylosLabini_etal_2020}. In particular, introducing an exponential cutoff for wavenumbers $k > k_c$ produces an approximately flat two-point correlation function on scales $r < r_c \approx 1/k_c$. In this situation, overdensities of characteristic size $r_c$ enter the non-linear regime nearly simultaneously, naturally triggering a monolithic, top-down collapse rather than hierarchical growth characteristic of power-law power-spectra.

The rapid temporal variation of the system's mean-field gravitational potential during this collapse generically leads to relaxed configurations in which dark and baryonic components acquire similar spatial and kinematic distributions. This provides a natural physical basis for the DMD phenomenology discussed above, in which the dominant mass component is distributed in a flattened, rotation-supported configuration traced by the \HI{} disk. Within this framework, the observed correlations between the kinematics of gas, stars, and the inferred dark component arise as a direct consequence of their shared dynamical origin.

This behavior stands in clear contrast to the standard CDM scenario, where, as discussed above,  power is present on all small scales and non-linear structures form hierarchically through a bottom-up sequence of mergers and accretion events, producing quasi-spherical dark matter halos that are dynamically decoupled from the baryonic disk. Accordingly, while hierarchical growth generically predicts cuspy inner density profiles, monolithic collapse scenarios naturally yield flat central cores \citep{Benhaiem+SylosLabini+Joyce_2019, SylosLabini_etal_2020}. 

This qualitative dynamical distinction underscores the importance of employing complementary diagnostics---such as those developed and applied in the present work---to discriminate between fundamentally different galaxy formation pathways.

}

%%%%%%%%%%%%%%%%%%%%%%%%%%%%%%%%%%%%%%%%%%%%%%%%%%%%%%%
%%%%%%%%%%%%%%%%%%%%%%%%%%%%%%%%%%%%%%%%%%%%%%%%%%%%%%%
%%%%%%%%%%%%%%%%%%%%%%%%%%%%%%%%%%%%%%%%%%%%%%%%%%%%%%%

\section{Conclusion} 
\label{sec:dis+con}

The aim of this work was to study the kinematic field of ESO~358-60 using the Velocity Ring Model (VRM), complemented by a diagnostic test specifically designed to assess the possible presence of a warp. Our main result is that the galaxy hosts a regular, unwarped disk, with a bar-like structure in its central region and significant radial motions in the outermost parts. From this analysis, we are able to reconstruct the two-dimensional coarse-grained radial and transverse velocity maps.  

These findings differ significantly from those of \citet{Kamphuis_etal_2025}, who, using a method based on the Tilted Ring Model (TRM), identified a pronounced warp of about \(20^\circ\) in the central region of the disk, together with strong radial motions in the same area. In their model, radial motions are negligible in the outskirts, in contrast with our results.  

These discrepancies likely arise from the different geometric assumptions of the two models: the VRM assumes a flat disk, whereas the TRM allows for geometric warps. However, observational evidence indicates that warps are generally found in the outer regions of edge-on disk galaxies \citep{Sancisi_1976, Reshetnikov_Combes_1999, Sanchez-Saavedra_etal_2003, Reshetnikov_etal_2016, Peters_etal_2017}, typically emerging near the optical radius, where tidal forces may become important. Moreover, observed warp angles rarely exceed \(10^\circ\) \citep{Garcia-Ruiz_etal_2002, Sanchez-Saavedra_etal_2003, Reshetnikov_etal_2016, Peters_etal_2017}.  

In contrast, the warp inferred by \citet{Kamphuis_etal_2025} appears unusually strong and centrally located. Our dedicated test, which combines elements of both the TRM and VRM approaches, suggests that this feature is more likely an artifact of the TRM assumptions. Additional support for this interpretation comes from the strong correlation between velocity dispersion and surface brightness, which indicates that the central density enhancement is better explained as a bar-like structure rather than an internal edge-on disk.  

The VRM analysis further reveals a rotationally supported disk extending out to about \( R_d \sim 150'' \). Within this radius, velocity anisotropies are small and consistent with regular circular rotation. Beyond this point, radial and tangential velocities become comparable (\( v_r \approx v_t \)), indicating a breakdown of purely circular motion. Any warp, if present, is therefore likely confined to the outermost regions.  

We have modeled the mass distribution of ESO~358-60 using two different approaches: the standard Navarro-Frenk-White (NFW) halo model and the Dark Matter Disk (DMD) model, in which the dark matter follows the distribution of neutral hydrogen. The DMD model, with a single free parameter, provides a better fit to the observed rotation curve than the two-parameter NFW model. Furthermore, it yields a total mass roughly four times smaller than that inferred from the halo-based model, and it is naturally compatible with a cored mass density profile, in agreement with the observed behavior of the rotation curve.

{ 
Recently, we investigated the velocity fields of dwarf galaxies drawn from the Local Irregulars That Trace Luminosity Extremes (LITTLE THINGS) survey and from the \HI{} Nearby Galaxy Survey (THINGS) \citep{SylosLabini_etal_2025a}, constructing two--dimensional velocity maps that simultaneously trace both the rotational and radial components of the flow.

In close analogy with the case of ESO~358--60, we found that within the radial domain where velocity anisotropies are sufficiently small for the disk to be considered rotationally supported, and where departures from planarity due to warps can be neglected, the DMD model yields fits that are statistically comparable to---and in several cases superior to---those obtained within the standard dark matter halo framework. A key quantitative difference between the two approaches is that the total masses inferred from the DMD model are typically lower by factors of $\sim 10$--$100$ relative to those derived in halo-based models.

A central feature of the DMD framework is that the inner slope of the rotation curve is directly determined by a linear combination of the surface density profiles of the stellar and gaseous components, which generally display flat cores. Consequently, the approximately linear rise of the circular velocity observed in the central regions of galactic disks arises naturally in this model, providing a simple and physically motivated explanation for a ubiquitous observational property. This stands in contrast to the halo paradigm, which often encounters difficulties in reproducing the observed inner kinematics without invoking fine-tuning or ad hoc modifications.

Consistent with these findings, earlier studies have identified several gas-rich yet dark-matter-dominated disk galaxies exhibiting similar behavior (e.g.\ DDO~154; \citealp{Carignan+Freeman_1988}). In such systems, improved fits to the rotation curves are obtained by scaling the \HI{} mass by a factor comparable to that found in our analysis (e.g.\ \citealp{Swaters_etal_2012}). Moreover, this type of scaling is also supported by studies of the baryonic Tully--Fisher relation, where it has been shown to provide a better description than alternative formulations, particularly when samples include low-mass, gas-rich galaxies \citep{Pfenniger+Revaz_2005}.

The fact that the DMD model appears statistically comparable to---if not better than---the NFW profile, particularly in the case of ESO~358--60, motivates a more detailed investigation of the dynamics of monolithic collapse in a cosmological context, as this mechanism may represent a viable alternative pathway for non-linear structure formation. We will explore this possibility in a forthcoming publication. As a concluding remark, we emphasize that our analysis of the kinematic field of ESO~358--60 pertains to a single system and does not yield broader insights into the general galaxy population. Therefore, our results do not alter the conclusions of \citet{Kamphuis_etal_2025} regarding the association between ESO~358--60 and the Fornax cluster.
}

%%%%%%%%%%%%%%%%%%%%%%%%%%%%%%%%%%%%%%%%%%%%%%%%%%%%%%%%%%%%%%

\begin{acknowledgements}

{  We are grateful to Douglas Altshuler for his insightful comments and valuable discussions.  
We acknowledge the use of data from the MeerKAT Fornax Survey, available at\\
 {\tt https://sites.google.com/inaf.it/meerkatfornaxsurvey/},  \\
and from the HyperLeda database, available at \\ 
{\tt http://leda.univ-lyon1.fr}.  
}
\end{acknowledgements}

% Please note that we have included the references below in
% order to compile the document, but we ask you to:
%   \bibliography{bibliography.bib} % your references Yourfile.bib

\end{document}